\documentclass[twoside,10pt]{article}
\usepackage{amssymb}
\usepackage{float}
\usepackage{amsmath}
\usepackage[utf8]{inputenc}
\usepackage[T1]{fontenc}
\usepackage{txfonts} 
\usepackage{color}
\usepackage{xspace}
\usepackage{graphicx}
\usepackage{authblk}
\usepackage{epstopdf}
\usepackage[centerlast, small,labelsep=period,textfont=normalsize,up]{caption} 
\usepackage{abstract} 
\usepackage{mathrsfs} 

\usepackage{format-rapport}
 
\graphicspath{{Figures/}}

\makeatletter{}
\newcommand{\DR}{D_t R}
\newcommand{\fe}[0]{f}
\newcommand{\vect}[1]{\mathbf{#1}}

\renewcommand{\phi}{\varphi}
\newcommand{\bPhi}{\boldsymbol{\varphi}}
\newcommand{\bXi}{\boldsymbol{\xi}}
\newcommand{\ba}{\boldsymbol{a}}
\newcommand{\bb}{\boldsymbol{b}}
\renewcommand{\epsilon}{\varepsilon}

\newcommand{\W}{\vect{W}}

\newcommand{\G}{\boldsymbol{G}}

\newcommand{\bV}{\mathbf{V}}

\newcommand{\demi}{\frac{1}{2}}

\newcommand{\ARonde}{\mathcal{A}}

\newcommand{\petitO}{\text{\normalfont o}}

\newcommand{\LRonde}{\mathcal{L}}

\newcommand{\ds}{\displaystyle}

\renewcommand{\div}{\text{div}}
\newcommand{\grad}{\boldsymbol{\nabla}}

\newcommand{\tw}{\textwidth}

\renewcommand{\emph}{\textit}

\def\bx{\mathbf{x}}
\def\bX{\mathbf{X}}
\def\bu{\mathbf{u}}

\newcommand{\bZero}{\boldsymbol{0}}

\newcommand{\cWood}{\text{c}_{\text{Wood}}}
\newcommand{\cFrozen}{\text{c}_{\text{Frozen}}}
\newcommand{\cPhase}[0]{\text{c}_{\text{Phase}}}

\newcommand{\kMuEpsilon}[0]{k^{\varepsilon,\microin}}
\newcommand{\kEpsilon}[0]{k^{\varepsilon}}
\newcommand{\cPhaseMuEpsilon}[0]{\text{c}^{\varepsilon,\microin}_{\text{Phase}}}
\newcommand{\cPhaseEpsilon}[0]{\text{c}^{\varepsilon}_{\text{Phase}}}
\newcommand{\betaMuEpsilon}[0]{\beta^{\varepsilon,\microin}}
\newcommand{\betaEpsilon}[0]{\beta^{\varepsilon}}
\newcommand{\resofreq}[0]{\omega_\text{res}}
\newcommand{\resofreqref}[0]{\omega_\text{res}^\text{ref}}
\newcommand{\extinctfreqref}[0]{\omega_\text{ext}^\text{ref}}
\newcommand{\betaref}[0]{\beta^\text{ref}}
\newcommand{\cPhaseref}[0]{\text{c}_\text{Phase}^\text{ref}}
\newcommand{\microin}[0]{\nu}

\newcommand{\epsilonlin}[0]{\epsilon_{\text{Lin}}}
\newcommand{\mulin}[0]{\microin_{\text{Lin}}}
\newcommand{\epsilonRP}[0]{\epsilon_{\text{RP}}}
\newcommand{\muRP}[0]{\microin_{\text{RP}}}

\definecolor{DarkGreen}{rgb}{0.,0.5,0.}
\definecolor{Bordeaux}{rgb}{0.75,0.,0.}
\definecolor{AppleGreen}{rgb}{0.,0.75,0.}
\definecolor{orange}{rgb}{0.97,0.65,0.02}
\definecolor{purple}{rgb}{0.61,0,0.99}
\definecolor{pink}{rgb}{0.98,0,0.99}

\newcommand{\beq}[1]{\begin{equation}\label{#1}}
\newcommand{\eeq}{\end{equation}}
\newcommand{\bitem}{\begin{itemize}}
\newcommand{\eitem}{\end{itemize}}

\title{A hierarchy of simple hyperbolic two-fluid models for bubbly flows}
\author[1,2,3]{Florence Drui}
\author[1,2,4]{Adam Larat}
\author[3,5]{Samuel Kokh}
\author[1,2,4]{Marc Massot}
\affil[1]{\normalsize \fontfamily{phv} {CNRS, UPR 288, Laboratoire d'Energ\'etique mol\'eculaire et macroscopique, combustion, Grande Voie des Vignes, 92295 Chatenay-Malabry, France}}
\affil[2]{\normalsize \fontfamily{phv} {Ecole Centrale Paris, Grande Voie des Vignes, 92295 Chatenay-Malabry, France}}
\affil[3]{\normalsize \fontfamily{phv} {Maison de la Simulation USR 3441, Digiteo Labs, b\^at. 565, PC 190, CEA Saclay, 91191 Gif-sur-Yvette, France.}}
\affil[4]{\normalsize \fontfamily{phv} {F\'ed\'eration de Math\'ematiques de l'Ecole Centrale Paris, FR CNRS 3487, France}}
\affil[5]{\normalsize \fontfamily{phv} {CEA/DEN/DANS/DM2S/STMF -  CEA Saclay, 91191 Gif-sur-Yvette, France.}}

\date{\normalsize \fontfamily{phv} {[Received date; Accepted date] - to be inserted later}}

\begin{document}
\maketitle
\begin{abstract}
\noindent \makeatletter{}With the objective of modeling both separate and disperse two-phase flows, we use in this paper
a methodology for deriving two-fluid models that do not assume any flow topology. 
This methodology is based on a variational principle and on entropy dissipation requirement. 
Some of the models that are such derived and studied are already known in the contexts of the description 
of separate- or disperse-phase flows. However, we here propose an arrangement of these models 
into a hierarchy based on their links through relaxation parameters. 
Moreover, the models are shown to be compatible with the description of a monodisperse bubbly flow 
and, within this frame, the relaxation parameters can be identified. 
This identification is finally verified and discussed through comparisons with experimental
measures of sound dispersion and with dispersion relations of a reference model for bubbly media.
 
\end{abstract}

\setcounter{tocdepth}{2}

\section{Introduction - Context}
Two-phase flows configurations may involve different regimes in the sense that the topology of the interface separating both fluids may involve very different scales
(see \citep{ishii} for examples of different flow regimes).
Indeed, in flows called \textit{separate-phase} flows, the variation of the material interface is described at the bulk fluid scale, while for bubbly flows or sprays, one phase is dispersed within a surrounding fluid and the interfaces are much smaller than the macroscopic dynamics of the flow.
Industrial applications may often involve flows where different regimes are at play successively or simultaneously.
For example in combustion chambers, as described in \citep{kah,chenadec}, a liquid fuel is injected at high velocity and pressure. In the vicinity of the injection duct, the fluids form a jet that can be considered as a separate-phase flow. However, further away from the injection point, the flow turns into a cloud of liquid drops. As there is a strong connection between the repartition of the droplets with respect to size and velocities and the quality of the combustion (see for instance a study in \citep{reveillon2005}), being able to accurately model the creation of this cloud in the chamber is a key challenge for the industrial community. Modeling such process is very difficult as it involves phenomena with very different characteristic scales, as is for instance widely described in \citep{letouze}.

The goal of the present work is to bring out connections between a system that may be used for modeling a dispersed flow, in the context of bubbly flows, and systems pertaining to the category of separate-phase flows models. Our approach leads to consider a five-equation barotropic two-phase model that is \textit{a priori} neutral with respect to the topology of the interfaces and brings into play a hyperbolic convective part supplemented with stiff source terms. A key feature of this model is that it accounts for two-scale kinematic effects: bulk kinematics and also small-scale vibrations \citep{gavrilyuk2002,gavrilyuk2011}. 
On the one hand, by considering instantaneous equilibria within the model, we will show that it can be considered as a parent model for two sub-models that are used in the literature for the simulation of separate-phase flows~\citep{ONERA-chante,grenier2013,champmartin2014}.
On the other hand, our five-equation model is compatible with hypotheses that describe the topology of a simple monodisperse bubbly flow. More specifically it is possible to retrieve an evolution equation for the bubble dynamics that is analogous to the Rayleigh-Plesset equation~\citep{rayleigh,plesset77}.
That way, we obtain a hierarchy of two-phase models that includes models suitable to both separate and disperse phase flows.
 Moreover, using this hierarchy, we propose two different estimates for two relaxation parameters related to mechanical equilibrium between materials in the separate-phase model. This result is notable as these estimates are often replaced by infinitely fast relaxation processes or heuristic values \citep{ONERA-chante,gallouet04,herard2005,saurel2009}. Few works have already addressed this matter and proposed similar estimate. However, the choice of this parameter must be careful as
an ill-considered estimate can lead to significantly underestimate damping effects in the system, as it will be shown in the sequel.
 The different choices for these fluid parameters will be tested by studying the acoustic regime and comparison with reference data issued from both experiments and the bubbly flows model of \citep{drew-passman}. This study will also shed some light on similarities and differences between the models of the hierarchy by considering their acoustic behavior as a benchmark tool.

The paper is structured as follows. First, we will apply classical modeling guidelines: the conservative part of the five-equation model will be derived thanks to the Least Action Principle. 
The system will be equipped with dissipative structures by enabling an entropy budget. After completing the definition of the parent five-equation model, two sub-models will be derived.
We shall then consider the case of a bubbly flow and identify the parent five-equation model with a monodisperse flow. In this context, we will see
that it is possible to identify parameters of the models with micro-viscosity and micro-inertia. Furthermore, estimate for these parameters will be proposed.
We shall then study each model of the hierarchy in the acoustic regime, by exploiting their respective dispersion relation. Finally, we will compare the acoustic behavior of the models with reference data.

\section{A five-equation hyperbolic two-fluid system}
\label{par: model derivation}

In the following section, we construct a complete model for the simulation of a two-phase flow within the framework of the following three ground hypotheses:
\begin{enumerate}
\item[(H1)] there is no mass transfer between the two phases, 
\item[(H2)] there is no shift velocity between the two phases, 
\item[(H3)] energy is a mute variable, completely defined by the partial densities and the local velocity. 
\end{enumerate}

In order to give a physical justification to these hypotheses, we illustrate them in the case of a bubbly flow. 
The limits of such hypotheses will be discussed in sections \ref{sec:Application} and \ref{par:measures}. Hypothesis (H1) 
is quite self-explanatory: there is no phase change, what is initially gaseous stays gaseous and the same for the liquid part. 
Hypothesis (H2) implies that the difference between the bulk velocity of both phase is neglected. This boils down to assume that velocity of the bubbles relaxes towards the velocity of the underlying carrying liquid with a characteristic time that is very small compared to other characteristic times that drive the flow. 
Consequently, we shall assume that a single velocity field $\bu$ can be used to describe the macroscopic kinematics of both phases.
Eventually, physical justification of hypothesis (H3) is more delicate. In \citep{prosperetti77}, Prosperetti studies thermal effects 
in bubble oscillations and proves that, in the small perturbation limit, the bubble oscillation mechanism is driven 
by three dimensionless numbers:
$$ 
  G_1 = \frac{\overline{\lambda}}{\lambda_g}, \quad 
  G_2 = \frac{R_0}{\varepsilon_{g}^{\text{th}}} 
  \quad \text{ and }\quad 
  G_3 = \frac{R_0}{\varepsilon_{l}^{\text{th}}}, 
$$
where $\bar{\lambda}$ is the mean free path in the gas, $\lambda_g$ is the wavelength within the bubble, 
$R_0$ is the bubble radius at equilibrium and $\varepsilon_{g}^{\text{th}}$ and $\varepsilon_{l}^{\text{th}}$ 
are the characteristic thicknesses of penetration of thermal conduction within the gas and the liquid. 
$G_1$ is shown to measure a ratio between acoustic and thermal phenomena characteristic penetration thicknesses 
and $G_1 \ll 1$ for a broad range of acoustic frequencies. Next, $G_2$ compares the thermal penetration characteristic 
thickness with the bubble radius and this allows us to discriminate two main regimes. When $G_2\ll 1$, thermal equilibrium is 
always reached within an acoustic period and the bubble oscillation can be considered as isothermal. When $G_2\gg 1$, thermal 
conduction between liquid and gas is negligible and the transformation can be considered as adiabatic. In both cases, internal 
energy is simply driven by the other state variables and the energy equation is redundant. 
Hypothesis (H3) thus extends this last statements to all the regimes in $G_2$, meaning that, with $G_2$, bubbles oscillation 
goes from an isothermal to an adiabatic regime under the constant fact that energy is always a mute variable. 

As a consequence, in the case of bubbly flows or not, both fluids are governed by a barotropic Equation Of State (EOS) characterized by 
\begin{equation}\label{eq:BarotropicEOS}
  \rho_k \mapsto {\fe}_k(\rho_k) 
  \quad \text{ and } \quad
  \rho_k \mapsto p_k(\rho_k) = \rho_k^2 (\text{d} \fe{}_k/\text{d}\rho_k)(\rho_k),
\end{equation}
where $\rho_k$, $\fe_k$ and $p_k$ are respectively the densities, the specific Helmoltz free energies and the partial pressures of each fluid, $k=1,2$. 
Then, $c_k^2=\left(\text{d} p_k/ \text{d}\rho_k\right)$ denote the sound velocities within each pure material $k$.
If we note $Y_k$ the mass fraction of each fluid, we have $Y_1 + Y_2 = 1$ and if we moreover postulate that fluids $1$ and $2$ are immiscible, 
we can also define the volume fractions $\alpha_k$, such that $\alpha_1 + \alpha_2 = 1$.
For convenience, in the rest of this paper we set $\alpha = \alpha_1$, $Y=Y_1$ and $m_k = \alpha_k \rho_k$, what allows us to define 
the density $\rho$ of the medium by
\[
\rho  = m_1 + m_2.
\]
Finally, we suppose given a function 
\[
(\rho,Y,\alpha)\mapsto \fe,
\]
that is the Helmoltz free energy of the medium. The choice of $\fe$ will be specified later.

Thanks to hypotheses (H1) and (H2), the total mass and the mass fractions of each fluid are conserved under the same velocity field $\bu$ and these hypotheses can be written as:
\begin{align}
\partial_t \rho + \div(\rho \bu)& =0,\label{eq:TotalMassConservation}
\\
\partial_t (\rho Y) + \div(\rho Y \bu)& =0,\label{eq:MassFractionsConservation}
\end{align}
or equivalently by $\partial_t m_k + \div(m_k \bu) =0$, $k=1,2$.

\subsection{Variational principle for two-phase models}
\label{par:variational-principle}

The first step in our modeling work aims at deriving a system of conservation laws, that is hyperbolic and equipped with a mathematical entropy evolution equation. Following the lines of 
\citep{berdichevsky} for barotropic fluids, of \citep{gavrilyuk2002} for a system of equations for two compressible fluids and two temperatures, or of \citep{caro} for a homogeneous
isothermal two-fluid model, we propose to use a variational approach 
and Hamilton's principle of stationary action to derive the conservative structure of our two-fluid model. 
Before going any further, we introduce a few notations: if $\ba$ and $\bb$ are two column-vectors whose components are $a_i$ and $b_i$, $1\leq i\leq d$, 
then $\ba^T$ is a row-vector, $\ba^T \bb = \ds\sum_{i} a_i b_i$ is a real number and $\ba \bb^T$ is a square matrix of size $d$ with $(\ba \bb^T)_{i,j} = a_i b_j$. 
If $\bx\in\mathbb{R}^d \mapsto A$ is a field of square matrices of size $d$ then $\div(A)$ is a column vector of size $d$, where $\div(A)_i = \sum_j \partial_j A_{ij}$. 
Finally, we note $D_t(\cdot)  = \partial_t(\cdot)+ \bu^T\boldsymbol{\grad}(\cdot)$ the material derivative.

Now, we need to define the kinetic energy $E_\text{kin}$ and the potential energy $E_\text{pot}$ that are involved in the conservative transformations of our system. 
We postulate that the specific potential energy of our system is the free energy $\fe$. 
Also, following \cite{gavrilyuk2002}, a key element of our study consists in considering a two-scale kinetic energy composed by a bulk kinetic energy $\rho|\bu|^2/2$ and a mesoscopic kinetic energy $\microin(D_t \alpha)^2/2$, where 
$\microin>0$ is a function of the volume fraction $\alpha$ only and whose physical interpretation and characterization is given latter on. 
This mesoscopic kinetic energy can be illustrated in the context of bubbly flows, as taking into account the vibrations of the bubbles: it models the bubbles as mesoscopic resonators by representing their pulsating 
inertia. Obviously, this split of the kinetic energy term is not specific to the context of bubbly flows. It allows a finer representation of general two-phase flows
and the consequences of such a hypothesis will be further discussed in sections \ref{sec:Application} and \ref{par:measures}.

Now, the Lagrangian $\LRonde = E_\text{kin} - E_\text{pot}$ of our system is fully defined, namely:
\begin{equation}
\label{eq:lagrangian-CVV-mu}
	\LRonde(\bu,\rho,Y,\alpha,D_t\alpha) =
	 \demi \rho |\bu|^2 + \demi \microin \left(D_t \alpha \right)^2 - \rho \fe(\rho, y, \alpha)
	 .
\end{equation}
Let $\mathscr{V}(t)$ be the volume occupied by a portion of fluid during $t_0\leq t \leq t_1$. We note
\(
\Omega=\{
(\bx,t)\in\mathbb{R}^3\times[t_0,t_1]~|~ \bx\in \mathscr{V}(t)
\}
\),
the subset of all space-time points generated by $\mathscr{V}(t)$ for $t_0\leq t\leq t_1$.
The space variable $\bX\in\mathscr{V}(0)$ denotes the Lagrangian coordinates associated with the reference frame at instant $t=t_0$. If $(\bX,t)\mapsto\bPhi$ is the mapping that gives the position $\bPhi(\bX,t)$ at instant $t$ of a fluid element that was located at $\bX$ at time $t=t_0$, then obviously $\Omega=\{ (\bPhi(\bX,t),t)~|~ \bX\in\mathscr{V}(t_0),\ t_0\leq t \leq t_1\}$.

From a pure Eulerian point of view, a transformation of the medium is fully characterized by the fields $(\bx,t)\mapsto (\rho,\bu,Y,\alpha)$. Equivalently, we can say that a transformation of the medium is fully characterized by the Eulerian fields $(\bx,t)\mapsto (Y,\alpha)$ and the Lagrangian mapping $(\bX,t)\mapsto\bPhi$ under the hypothesis that $\bPhi$ is compliant with the mass conservation.

If $(\bx,t)\mapsto (Y,\alpha)$ and $(\bX,t)\mapsto\bPhi$ is a given transformation of the medium, we consider a family of transformations $(\bx,t,\zeta)\mapsto (\widehat{Y},\widehat{\alpha})$ and
$(\bX,t,\zeta)\mapsto\widehat{\bPhi}$ parametrized by $\zeta\in[0,1]$ such that:
\begin{itemize}
\item $(\widehat{Y},\widehat{\alpha})(\bx,t,\zeta=0) = (Y,\alpha)(\bx,t)\;$ and 
$\; \widehat{\bPhi}(\bX,t,\zeta=0) = \bPhi(\bX,t)$,
\item
 $\widehat{Y}$ and $\widehat{\bPhi}$ verify the mass and partial mass conservation \eqref{eq:TotalMassConservation} and 
 \eqref{eq:MassFractionsConservation}, for all $\zeta\in[0,1]$.
\end{itemize}
We adopt the classic definition of the infinitesimal transformations that acts on the medium by introducing the infinitesimal displacement $(\bx,t)\mapsto\bXi$, where
$$
\bXi\big(\bPhi(\bX,t),t\big) = 
\left(
\frac{\partial \widehat{\bPhi}}{\partial\zeta}
\right)_{\bX,t}(\bX,t,\zeta=0),
$$
and by setting for any Eulerian field $(\bx,t,\zeta)\mapsto \widehat{b}$
$$
\delta b (\bx,t)
=
\left(
\frac{\partial \widehat{b}}{\partial \zeta}
\right)_{\bx,t}(\bx,t,\zeta=0).
$$
Using the lines of \cite{berdichevsky,gavrilyuk2002,gavrilyuk2011}, our hypotheses provide the infinitesimal variations of $\rho$, $Y$, $u$ and $D_t \alpha$. We obtain indeed that
\begin{align}
\delta \rho &= -\div(\rho \bXi) ,&
\delta \bu &= D_t \bXi - (\bXi^T \grad) \bu,&
\delta Y &= - (\bXi^T \grad) Y 
\quad \text{ and }\quad 
\delta(D_t\alpha) &= D_t(\delta\alpha) 
+ (\grad\alpha)^T [D_t\bXi - \div( \bu \bXi^T)] 
.
\end{align}

Now, we can define the Hamiltonian Action $\ARonde(\zeta)$ by setting
$$
\ARonde(\zeta) 
= \int_{\Omega} 
\LRonde(\widehat\bu,\widehat\rho,\widehat Y,\widehat \alpha, 
\partial_t\widehat{\alpha} + \widehat{\bu}^T\grad\widehat{\alpha})
\,\text{d}\bx\text{d}t,
$$
and compute the infinitesimal variations of the action, $\delta \ARonde = (\text{d} \ARonde / \text{d} \zeta) (\zeta = 0)$: 
$$
\delta \ARonde
= \ds \int_{\Omega} 
\left[
\left(\frac{\partial\LRonde}{\partial \bu}\right)^T
\delta \bu
+
\frac{\partial\LRonde}{\partial\rho}
\delta \rho
+
\frac{\partial\LRonde}{\partial Y}
\delta Y
+
\frac{\partial\LRonde}{\partial\alpha}
\delta \alpha
+
\frac{\partial\LRonde}{\partial(D_t \alpha)}
\delta (D_t \alpha)
\right]
(\bu,\rho, Y, \alpha,  D_t \alpha)
\,\text{d}\bx\text{d}t.
$$
We make the classic assumption (see \cite{gavrilyuk1999,gavrilyuk2002}) that for our transformation family, 
$\bXi$ and $\delta\alpha$ vanish on $\partial\Omega$. Thanks to the Green formula, 
we obtain after tedious calculations that
\begin{multline}
\delta \ARonde = 
-\int_{\Omega}
\left\{
- \rho \grad\left(\frac{\partial \LRonde}{\partial \rho}\right)
+ \partial_t\left(\frac{\partial \LRonde}{\partial \bu}\right)
+ \div\left(\frac{\partial \LRonde}{\partial \bu} \bu^T\right)
+ (\grad\bu)^T\left(\frac{\partial \LRonde}{\partial \bu}\right) 
+ \left(\frac{\partial \LRonde}{\partial Y}\right) \grad Y
\right.
\\
\left.
+ \left[
\partial_t\left( \frac{\partial \LRonde}{\partial(D_t\alpha)}\right)
+
\div\left( \frac{\partial \LRonde}{\partial(D_t\alpha)} \bu\right)
\right]
\grad\alpha
+ 
 \frac{\partial \LRonde}{\partial(D_t\alpha)}
 \grad(D_t\alpha)
\right\}^T\bXi
\,\text{d}\bx\text{d}t
\\
-\int_{\Omega}
\left[
-\frac{\partial\LRonde}{\partial \alpha}
+\partial_t\left(
\frac{\partial\LRonde}{\partial (D_t\alpha)}
\right)
+\div\left(
\frac{\partial\LRonde}{\partial (D_t\alpha)}
\bu
\right)
\right]
\delta\alpha
\,\text{d}\bx\text{d}t
.
\end{multline}
We follow the Least Action Principle that boils down to postulate that a physical transformation of the medium should extremize the Hamiltonian action $\ARonde$. In our case, this yields
\begin{align}
- \rho \grad\left(\frac{\partial \LRonde}{\partial \rho}\right)
+ \partial_t\left(\frac{\partial \LRonde}{\partial \bu}\right)
&+ \div\left(\frac{\partial \LRonde}{\partial \bu} \bu^T\right)
+ (\grad\bu)^T\left(\frac{\partial \LRonde}{\partial \bu}\right) 
+ \left(\frac{\partial \LRonde}{\partial Y}\right) \grad Y
\notag
\\
&+ \left[
\partial_t\left( \frac{\partial \LRonde}{\partial(D_t\alpha)}\right)
+
\div\left( \frac{\partial \LRonde}{\partial(D_t\alpha)}\bu\right)
\right]
\grad\alpha
+ 
 \frac{\partial \LRonde}{\partial(D_t\alpha)}
 \grad(D_t\alpha)
&=\bZero,
\label{eq: least action eq for momentum}
\\
&-\frac{\partial\LRonde}{\partial \alpha}
+\partial_t\left(
\frac{\partial\LRonde}{\partial (D_t\alpha)}
\right)
+\div\left(
\frac{\partial\LRonde}{\partial (D_t\alpha)}
\bu
\right)
&=0
.
\label{eq: least action eq for alpha}
\end{align}

For the choice of $\LRonde$ expressed by \eqref{eq:lagrangian-CVV-mu}, we have
\begin{align}
\frac{\partial\LRonde}{\partial\rho}
&=\frac{|\bu|^2}{2} - \fe{} - \rho\frac{\partial \fe{}}{\partial\rho}
,&
\frac{\partial\LRonde}{\partial Y}
&=-\rho\frac{\partial \fe{}}{\partial Y}
,&
\frac{\partial\LRonde}{\partial \alpha}
&=\demi  \microin'(\alpha) \left( D_t \alpha \right)^2
-\rho\frac{\partial \fe{}}{\partial \alpha} 
,&
\frac{\partial\LRonde}{\partial (D_t\alpha)}
&=\microin D_t\alpha
,&
\frac{\partial\LRonde}{\partial \bu}
&=\rho\bu
.
\label{eq: partial derivative of L}
\end{align}
Relations~\eqref{eq: least action eq for momentum} and \eqref{eq: least action eq for alpha} will respectively provide the evolution equations for the momentum and the volume fraction. Indeed, reinjecting \eqref{eq: partial derivative of L} into \eqref{eq: least action eq for momentum}-\eqref{eq: least action eq for alpha} provides
\begin{align*}
\partial_t(\rho\bu)
+\div(\rho \bu\bu^T)
+\grad\left(
\rho^2\frac{\partial \fe{}}{\partial \rho}
+\frac{1}{2}\microin (D_t\alpha)^2
\right)
&=\bZero
,&
- \rho \frac{\partial \fe{}}{\partial \alpha} 
& =  -\demi (D_t\microin) (D_t \alpha ) + \partial_t ( \microin D_t\alpha)  + \div(\bu \microin D_t\alpha )
.
\end{align*}
We define the medium pressure $p$ and a new variable $w$ by setting
\begin{align}
p=\rho^2\frac{\partial \fe{}}{\partial \rho}
\quad\text{ and }\quad
D_t \alpha = \frac{\rho Y w}{\sqrt{\microin}}
.
\label{eq: def of p and w}
\end{align}
Using the mass and partial mass conservation hypotheses we obtain that the fluid transformations are governed by the following system of equations

\begin{subequations}
\begin{align}
\partial_t \rho + \div(\rho \bu)
&=0,
\label{eq: system eq rho}
\\
\partial_t (\rho Y) + \div(\rho Y \bu)
&=0,
\label{eq: system eq rho Y}
\\
\partial_t(\rho\bu)
+\div(\rho \bu\bu^T)
+\grad\left(
p
+\frac{1}{2}\microin (D_t\alpha)^2
\right)
&=\bZero,
\label{eq: system eq rho u}
\\
\partial_t \alpha + \bu^T\grad \alpha &= \frac{m_1 w}{\sqrt{\microin}} ,
\label{eq: system eq alpha}
\\
\partial_t w + \bu^T\grad w &= - \frac{\rho}{m_1 \sqrt{\microin}} \frac{\partial \fe{}}{\partial \alpha}.
\label{eq: system eq w}
\end{align}
\label{eq: general consv system}
\end{subequations}

The Least Action Principle only provides here conservative elements for our model: the momentum equation~\eqref{eq: system eq rho u}, the evolution equation for the volume fraction~\eqref{eq: system eq alpha} and small scale pulsation evolution equation~\eqref{eq: system eq w}, that supplement the mass conservation equations~\eqref{eq: system eq rho}-\eqref{eq: system eq rho Y} which have been postulated. 
We shall examine the dissipative structures of system~\eqref{eq: general consv system} in the next section.

\subsection{Dissipation and second principle of thermodynamics}
\label{par:entropy-principle}

While we are not interested in bulk dissipation phenomena, we aim at describing small scale dissipation associated with the mesoscopic kinetic energy $\microin\left(D_t\alpha\right)^2/2$, see equation \eqref{eq:lagrangian-CVV-mu}.
This way, the damping of  local microscopic pulsations of the interface due to various dissipative phenomena can be 
taken into account. 

In the specific case of bubbly flows, we illustrate our words by the large study done in \citep{prosperetti77}, where 
the different damping contributions to bubbles oscillations are identified to be viscous, thermal and acoustic.
For very small bubbles (radius below $10^{-6}$ m), the dominant damping effect is the mechanical one, 
due to the viscous stress at the gas-liquid interface.
Thermal damping effects are dominant for bubbles of larger radii and at low frequency pulsations. This damping
is due to the heat exchanges between the gas and the liquid phases and to the thermodynamical state of the
interior of the bubble.  

Another form of damping is related to the “sound” emitted by the interface (or the bubbles) when it vibrates.
This phenomenon prevails for high frequencies perturbations. However, in this range of frequencies,
we believe that our model is not relevant, as it does not take into account diffraction, for instance.

In a general two-phase flow context, we now propose to introduce irreversible damping effects in our two-phase system, 
by adding terms to \eqref{eq: general consv system} that are compatible with a mathematical entropy evolution equation.
First, we suppose that equations \eqref{eq: system eq rho}-\eqref{eq: system eq alpha} are valid, but we discard \eqref{eq: system eq w} and consider that $D_t w$ is now a quantity to be defined.

In the barotropic case, it is classic (see \citep{godlewski,serre-fr}) to consider the total free energy of the system
as a mathematical entropy: 
\begin{equation}
\rho \eta (\bu, \rho, Y,\alpha, w) = \demi \rho |\bu|^2 + \demi
	\left(\rho Y w \right)^2 + \rho \fe{}(\rho,Y,\alpha).
\label{eq:entropy-def}
\end{equation}
We now seek for an entropy flux function $\G$ and a proper evolution principle for $D_t w$, such that 
$$
\partial_t(\rho\eta) + \div(\rho\eta\bu +\G) \leq 0,
$$
or equivalently
\begin{equation}
\rho D_t \eta  + \div(\G) \leq 0.
\label{eq: entropy inequality}
\end{equation}
Relation~\eqref{eq: entropy inequality} reads
\begin{equation*}
\div(\G) 
+ \rho\left(\frac{\partial \eta}{\partial\bu}\right)^T D_t\bu
+ \rho\frac{\partial \eta}{\partial\rho} D_t\rho
+ \rho\frac{\partial \eta}{\partial Y} D_t Y
+ \rho\frac{\partial \eta}{\partial\alpha} D_t\alpha
+ \rho\frac{\partial \eta}{\partial w} D_t w
\leq 0
.
\end{equation*}
Using \eqref{eq: system eq rho}-\eqref{eq: system eq alpha} to express 
$D_t\bu$, $D_t \rho$, $D_t Y$, $D_t\alpha$ and
\begin{align*}
\frac{\partial\eta}{\partial \bu} 
&= \bu
,&
\frac{\partial\eta}{\partial \alpha} 
&= \frac{\partial \fe{}}{\partial \alpha}
,&
\frac{\partial\eta}{\partial \rho} 
&=  \frac{1}{2} (Yw)^2 + \frac{\partial \fe{}}{\partial \rho}
,&
\frac{\partial\eta}{\partial w} 
&= \rho Y^2 w
,&
\end{align*}
altogether with \eqref{eq: def of p and w},  we obtain that
\begin{equation}
\div
\left[
\G - \left(
p + \frac{1}{2} (\rho Y w)^2
\right)\bu
\right]
+\frac{\rho^2 Y w}{\sqrt{\microin}} 
\left( 
\frac{\partial \fe{}}{\partial\alpha} + \sqrt{\microin} Y D_t w
\right)
\leq 0
.
\label{eq: entropy inequality intermediate}
\end{equation}
A simple choice for ensuring \eqref{eq: entropy inequality intermediate} consists in setting
\begin{align*}
\G = 
\left(
p + \frac{1}{2}  (\rho Y w)^2
\right)\bu
\quad \text{ and } \quad
 \frac{\partial \fe{}}{\partial\alpha} + \sqrt{\microin} Y D_t w
 = - \varepsilon \frac{Y w}{\sqrt{\microin}} 
,
\end{align*}
where $\varepsilon>0$ is a constant. This yields a definition of $\G$ 
and a new evolution equation for $w$ that reads
$$
\partial_t w + \bu^T\grad w = 
-\frac{\varepsilon}{\microin} w - \frac{1}{\sqrt{\microin} Y }\frac{\partial \fe{}}{\partial \alpha}
.
$$
Consequently, the generic form of our two-phase flow system reads
\begin{subequations}
\begin{align}
\partial_t \rho + \div(\rho \bu)
&=0,
\label{eq: general system + dissip eq rho}
\\
\partial_t (\rho Y) + \div(\rho Y \bu)
&=0,
\label{eq: general system + dissip eq rho Y}
\\
\partial_t(\rho\bu)
+\div(\rho \bu\bu^T)
+\grad\left(
p
+\frac{1}{2} (\rho Y w)^2
\right)
&=\bZero,
\label{eq: general system + dissip eq rho u}
\\
\partial_t \alpha + \bu^T\grad \alpha &= \frac{m_1 w}{\sqrt{\microin}} ,
\label{eq: general system +dissip eq alpha}
\\
\partial_t w + \bu^T\grad w &= 
-\frac{\varepsilon}{\microin} w 
- \frac{\rho}{m_1 \sqrt{\microin}} \frac{\partial \fe{}}{\partial \alpha}.
\label{eq: general system + dissip eq w}
\end{align}
\label{eq: general consv system + dissipation}
\end{subequations}

In order to complete the definition of our model, we need to specify the free energy of the medium. We consider that $\fe$ is the sum of a bulk mixture free energy and a compaction energy $\alpha\mapsto e(\alpha)$, where $e$ is a given function (see \cite{gavrilyuk2002}). We set
\begin{equation}
\fe{}(\rho,Y,\alpha) = 
Y \fe{}_1 \left(\frac{\rho Y }{\alpha}\right) 
+ (1-Y) \fe{}_2 \left(\frac{\rho (1-Y) }{1-\alpha}\right)
+ e(\alpha)
.
\end{equation}
For this choice, granted that $\rho_k^2 \partial\fe{}_k / \partial \rho_k = p_k$, $k=1,2$, a straightforward calculation gives that
\begin{align*}
p=\rho^2\frac{\partial\fe{}}{\partial \rho} = \alpha p_1 + (1-\alpha)p_2
\quad \text{ and } \quad
\rho \frac{\partial\fe{}}{\partial \alpha} &= p_2 - p_1 + \rho \frac{\text{d} e}{\text{d} \alpha},
\end{align*}
and system \eqref{eq: general consv system + dissipation} reads here
\begin{subequations}
\begin{align}
\partial_t \rho + \div(\rho \bu)
&=0,
\label{eq: system + dissip eq rho}
\\
\partial_t (\rho Y) + \div(\rho Y \bu)
&=0,
\label{eq: system + dissip eq rho Y}
\\
\partial_t(\rho\bu)
+\div(\rho \bu\bu^T)
+\grad\left(
p
+\frac{1}{2}\microin (D_t\alpha)^2
\right)
&=\bZero,
\label{eq: system +dissip eq rho u}
\\
\partial_t \alpha + \bu^T\grad \alpha &= \frac{m_1 w}{\sqrt{\microin}} ,
\label{eq: system +dissip eq alpha}
\\
\partial_t w + \bu^T\grad w &= 
-\frac{\varepsilon}{\microin} w 
+ \inv{m_1\sqrt{ \microin}} 
\left(
p_1 - p_2 - \rho\frac{\text{d} e}{\text{d}\alpha}
\right).
\label{eq: system +dissip eq w}
\end{align}
\label{eq: consv system + dissipation}
\end{subequations}
The new variable $w$ accounts for variations due to small scale velocities. 
It is worth noting these vibration-like effects
impact the total momentum of the mixture and appear as an additional pressure in the third equation of system \eqref{eq: general consv system + dissipation}.

Let us conclude this section by stating well-posedness properties of our two-phase  model with micro-inertia $\microin$. 
We consider the sole convective part of system~\eqref{eq: general consv system + dissipation} for one-dimensional problems by discarding the source terms. The resulting system is hyperbolic and its characteristic velocities are
$$
u-c, \quad u,\quad u+c,
$$
where
\begin{equation}\label{eq:cFrozen}
c^2 
= 
\cFrozen^2
+
\rho (Y w)^2
\quad \text{ and } \quad
\cFrozen^2 = 
Y c_1^2 + (1-Y) c_2^2
.
\end{equation}
Details related to the eigenstructure of system~\eqref{eq: general consv system + dissipation} are presented in appendix~\ref{section: appendix eigenstructure}.

\subsection{Submodels and traversing the hierarchy}
\label{par:hierarchy}

We consider the most complete model \eqref{eq: consv system + dissipation} and we examine two limit flow regimes, 
obtained for vanishing values of the parameters $\varepsilon$ and $\microin$. 
First, we study the case of a negligible micro-inertia compared to the internal dissipation effects, \textit{i.e.} $\microin\to 0$ and $\varepsilon = O(1)$. 
Second, we consider the case when both micro-inertia and internal dissipation tend to zero, with $\microin\to 0$ and $\varepsilon = \petitO(\sqrt{\microin})$. 
Both cases allow to recover the two-phase systems presented in \cite{ONERA-chante}. These systems are both composed of a conservative part and a (possibly null) stiff source term.

\subsubsection{4-equation model for $\microin\to 0$ and $\varepsilon = O(1)$}

We consider system~\eqref{eq: consv system + dissipation} and suppose here that $\microin\to 0$ for a fixed value of $\varepsilon$. We see that \eqref{eq: system +dissip eq w} provides that
$$
\frac{m_1 w}{\sqrt{\microin}} = \frac{1}{\varepsilon}
\left( 
p_1 - p_2 - \rho  \frac{\text{d} e}{\text{d} \alpha}
\right)
.
$$
By using equation~\eqref{eq: system +dissip eq alpha} we obtain that the limit regime is governed by
\begin{subequations}
\begin{align}
\partial_t \rho + \div(\rho \bu)
&=0,
\label{eq: cvv-epsilon rho}
\\
\partial_t (\rho Y) + \div(\rho Y \bu)
&=0,
\label{eq: cvv-epsilon rho Y}
\\
\partial_t(\rho\bu)
+\div(\rho \bu\bu^T)
+\grad p
&=\bZero,
\label{eq: cvv-epsilon rho u}
\\
\partial_t \alpha + \bu^T\grad \alpha &= 
\inv{\varepsilon} 
\left(
p_1 - p_2 - \rho\frac{\text{d} e}{\text{d}\alpha}
\right).
\label{eq: cvv-epsilon alpha}
\end{align}
\label{eq: cvv-epsilon}
\end{subequations}
In the specific case ${\text{d} e}/{\text{d}\alpha} = 0$, we 
recover the relaxation system studied in \cite{ONERA-chante}. The conservative part of system~\eqref{eq: cvv-epsilon} is a hyperbolic system whose characteristic velocities are $\{u -  \cFrozen, u, u + \cFrozen\}$ and that is equipped with an entropy inequality
$$
\partial_t
\left(
\frac{1}{2}\rho |\bu|^2 + \rho  g(\rho, Y,\alpha)
\right)
+\div
\left(
\left[
\frac{1}{2}\rho |\bu|^2 + \rho  g(\rho, Y,\alpha)
+p
\right]\bu
\right)
\leq 0
.
$$
The stiff source term in \eqref{eq: cvv-epsilon alpha} drives all the dissipation effects within system~\eqref{eq: cvv-epsilon}.

\subsubsection{3-equation model for $\microin\to 0$ and $\varepsilon = \petitO(\sqrt{\microin})$}
\label{par:3-equation}

The last model of our hierarchy can be obtained either by considering 
model~\eqref{eq: consv system + dissipation} and the two vanishing coefficients 
$\microin\to 0$ and $\varepsilon = \petitO(\sqrt{\microin})$, 
or by taking $\varepsilon\to 0$ in \eqref{eq: cvv-epsilon}. 
Formally, we  obtain
\begin{subequations}
\begin{align}
\partial_t \rho + \div(\rho \bu)
&=0,
\label{eq: CVV rho}
\\
\partial_t (\rho Y) + \div(\rho Y \bu)
&=0,
\label{eq: CVV rho Y}
\\
\partial_t(\rho\bu)
+\div(\rho \bu\bu^T)
+\grad p
&=\bZero,
\label{eq: CVV rho u}
\\
p_1\left(\frac{\rho Y}{\alpha}\right) 
- 
p_2\left(\frac{\rho (1-Y)}{1-\alpha}\right) 
- \rho\frac{\text{d} e}{\text{d}\alpha}(\alpha) &=0.
\label{eq: CVV alpha closure}
\end{align}
\label{eq: CVV}
\end{subequations}
This system is fully conservative. Also, relation~\eqref{eq: CVV alpha closure} implies that $\alpha$ is no longer an independent variable: the volume fraction has become a function of $\rho$ and $Y$.

In the specific case $\text{d} e/{\text{d}\alpha} = 0$, we recover the classic partial pressure equilibrium closure relation $p_1 = p_2 $ that was studied in \cite{ONERA-chante}: 
the resulting system is strictly hyperbolic as 
it possesses three distinct  real-valued characteristic velocities $\{u - \cWood, u , u+\cWood\}$, where $\cWood$ is defined by 
\begin{equation}
\label{eq:cWood}
	\frac{1}{\cWood^2} 
	= 
	 \frac{\alpha^2}{Y c_1^2} 
	 + 
	 \frac{(1-\alpha)^2}{ (1-Y) c_2^2}
	 .
\end{equation}

Concerning the mathematical properties of \eqref{eq: CVV}, we refer the reader to \citep{ONERA-chante}.
One can just note that for both relations \eqref{eq:cFrozen} and \eqref{eq:cWood}, the subcharacteristic condition is 
verified, as $\cWood \leq \cFrozen \leq c$. However, we will see in part \ref{par:dispersion relations} that
the real velocities of wave propagation are different from these characteristic velocities and depend on the wave frequency.

\section{Application to bubbly flows in the small perturbation regime} 
\label{sec:Application}

The derivation of  model~\eqref{eq: general consv system + dissipation} relies on general mechanics and thermodynamics principles that do not involve specific hypotheses on the flow topology. 
Hereafter, we show that under simple hypotheses, we obtain a specific model for bubbly flows, compatible with system 
\eqref{eq: consv system + dissipation}.
This allows us to carry out an acoustic regime analysis and so to compare the behaviour of our model with other bubbly flow models.

\subsection{Connection with the Rayleigh-Plesset equation}
\label{par:RP-comparison}

Let us consider a bubbly flow which is monodisperse and characterized as follows:
at each position and instant $(\bx,t)$, the distribution of the number of bubbles is defined by the density function $n(\bx,t)$ and all bubbles are spherical with a radius $R(\bx,t)$. Given the phasic gas density $\rho_1(\bx,t)$, all the bubbles  have the same mass $\mathcal{M}_b(\bx,t) = 4/3 \pi \rho_1(\bx,t) R^3(\bx,t)$. 
The gas volume fraction $\alpha$ and partial mass $m_1$ can now be related to the flow structure parameters by
\begin{equation}
\alpha(\bx,t) = \frac{4\pi }{3} R^3(\bx,t) n(\bx,t)
,\quad
m_1(\bx,t) = n(\bx,t) \mathcal{M}_b(x,t)
.
\label{eq: bubbly flow alpha and m_1}
\end{equation}

Now, we make two additional assumptions:
\begin{enumerate}
\item[(H4)] the mass of each bubble remains constant during a medium transformation (no break-up nor collapse),
\item[(H5)] the surrounding liquid is incompressible and thus has a constant density $\rho_2 = \overline{\rho}_2$.
\end{enumerate}

Hypothesis~(H4) boils down to $D_t \mathcal{M}_b =0$. Using the conservation  of the partial mass $m_1 =\rho Y$ for the gas \eqref{eq: system + dissip eq rho Y} and $n=m_1/\mathcal{M}_b$, we obtain the following conservation law for $n$
\begin{equation}
\partial_t n + \div (n \bu) = 0.
\label{eq: eq n}
\end{equation}
The conservation  of the partial mass $m_2=\rho (1-Y)$ for the surrounding fluid and hypothesis (H5) 
imply that $\div(\bu)$ is constrained by $D_t \alpha$ through the relation
\begin{equation}
D_t \alpha + (\alpha-1) \div (\bu) = 0.
\label{eq: constraint D_t alpha + u_x}
\end{equation}
Now, let us express $D_t \alpha$, $w$ and $D_t w$ in terms of $R(t)$ and its material derivatives. We have $\alpha = (4\pi/3) R^3 m_1/\mathcal{M}_b$, which yields
\begin{equation}
D_t \alpha = 
\alpha\left(- \div (\bu) + 3\frac{\DR}{R} \right)
.
\label{eq: D_t alpha expression}
\end{equation}
Using \eqref{eq: constraint D_t alpha + u_x}, we obtain
$$
D_t \alpha = 3 \alpha (1-\alpha) \frac{\DR}{R} 
.
$$
Expressing $m_1$ thanks to \eqref{eq: bubbly flow alpha and m_1} and using 
\eqref{eq: system +dissip eq alpha}
leads to
\begin{equation}
w =  \sqrt{\microin} \frac{4 \pi}{\mathcal{M}_b} (1-\alpha) R^2 \DR.
\label{eq: w expression}
\end{equation}
Relations~\eqref{eq: w expression} and \eqref{eq: D_t alpha expression} provide
\begin{equation}
D_t w 
= 
\frac{3 \alpha(1-\alpha)\sqrt{\microin}}{m_1}
\left[
\left(
2 - 3\alpha
+
\frac{3\alpha(1-\alpha)}{2}
\frac{\microin'}{\microin}
\right)
\left(
\frac{\DR}{R}
\right)^2
+
\frac{D_{tt}^2 R}{R}
\right]
.
\label{eq: D_t w expression}
\end{equation}
Finally, combining \eqref{eq: D_t w expression} and \eqref{eq: system +dissip eq w}  gives the evolution equation for $R$:
\begin{equation}
p_1 - p_2 - \rho\frac{\text{d} e}{\text{d}\alpha}
 = \varepsilon \frac{3 \alpha (1-\alpha)}{R} \DR 
\, + \, 3 \alpha (1-\alpha) \microin 
\left(3(1-\alpha) + \frac{3}{2} \alpha (1-\alpha) \frac{\microin'(\alpha)}{\microin}  \right) \left(\frac{\DR}{R}\right)^2
\, + \, 3 \alpha (1-\alpha) \microin \frac{D_{tt}^2 R}{R}
.
\label{eq: equation for R}
\end{equation}
We see that the two-phase model~\eqref{eq: consv system + dissipation}, supplemented by our bubbly flow structure assumptions (H4-H5), 
provide the evolution equation~\eqref{eq: equation for R} for a spatial distribution of bubbles radii. 

If we moreover suppose that: 
\begin{itemize}
\item[(H6)] for a bubble located at $(\bx,t)$, the pressure is uniform within the bubble and equal to $p_1(\bx,t)$ 
and the pressure of the surrounding liquid is equal to $p_2(\bx,t)$,
\item[(H7)] the radius distribution is uniform in space, namely $R=R(t)$,
\end{itemize}
then $R$ is a global variable of the studied multiphase system, whose dynamics is driven by relation~\eqref{eq: equation for R}, 
where $\DR = (\text{d}R /\text{d} t)(t) = \dot{R}(t)$ and $D_{tt}^2 R = (\text{d}^2R /\text{d} t^2)(t) = \ddot{R}(t)$.
This equation is now an ODE, analogous to 
the evolution equation of a nonlinear oscillatory system with damping and forcing terms
which reads
\begin{equation}
p_1 - p_2 - \rho\frac{\text{d} e}{\text{d}\alpha}
 = \varepsilon \frac{3 \alpha (1-\alpha)}{R} \dot{R}
\, + \, 3 \alpha (1-\alpha) \microin 
\left(3(1-\alpha) + \frac{3}{2} \alpha (1-\alpha) \frac{\microin'(\alpha)}{\microin}  \right) \left(\frac{\dot{R}}{R}\right)^2
\, + \, 3 \alpha (1-\alpha) \microin \frac{\ddot{R}}{R}
.
\label{eq: nonlinear ODE for R}
\end{equation}
Thanks to this analogy, $\microin$ is connected to the inertial effects and referred to as "micro-inertia", 
while $\varepsilon$ is related to damping.
Moreover, under the strong hypothesis (H7), one can compare the evolution equation of $R(t)$ 
with other existing models that account for bubble vibrations in specific flow regimes. 
These comparisons may provide an estimate for the values of $\microin$ and $\varepsilon$ in the model~\eqref{eq: consv system + dissipation}. We propose to proceed along these lines by considering the Rayleigh-Plesset equation~\eqref{eq: DP system - R} of the Drew-Passman system~\eqref{eq:DP-system}. 
We examine a flow regime involving a bubble radius and a volume fraction small enough such that $\alpha \ll 1$ in \eqref{eq: consv system + dissipation} and $kR \ll 1$ in \eqref{eq:DP-system}. We neglect the surface tension in \eqref{eq: DP system - R} by setting $\sigma = 0$ and relate
$p_1 - p_2 - \rho{\text{d} e}/{\text{d}\alpha}$ 
in \eqref{eq: equation for R} to 
$p_{1i} - p_2$ in  \eqref{eq: DP system - R}, recalling the hypothesis of uniform pressures (H6).
By identifying the terms in 
$\dot{R}$, $\ddot{R}$ and $\dot{R}^2$, we respectively obtain:
\begin{equation}
\varepsilon \frac{3 \alpha }{R}   = \frac{4 \mu_2}{R}
,\quad
\frac{3 \alpha}{R} \microin   = \rho_2 R
\quad\text{ and }\quad
\frac{3 \alpha}{R^2} \microin 
\left(\demi \microin^{-1} \microin'(\alpha) 3 \alpha 
 +  2 \right) = \frac{3}{2} \rho_2
.
\label{eq: comparison radius eq}
\end{equation}
The first two relations of \eqref{eq: comparison radius eq} allow to identify respectively $\varepsilon$ and $\microin$ as 
\begin{equation}
	\epsilonRP = \frac{4\mu_2}{3\alpha}
	\quad\text{ and }\quad
	\muRP = \frac{\rho_2 R^2}{3 \alpha }
	=
	\frac{\rho_2 (3\alpha)^{-1/3}}{(4 \pi n)^{2/3}}
	.
	\label{eq: evaluation of mu and epsilon}
\end{equation}
The third relation of \eqref{eq: comparison radius eq} is redundant but compatible with the definition of $\microin$ expressed in \eqref{eq: evaluation of mu and epsilon}. 
We thus see that for this specific flow regime, it is possible to insert a monodisperse bubble flow model into our two-phase model,
for which the dynamics of the bubble radii degenerates to the Rayleigh-Plesset equation. 
Also, the resulting values of $\microin$ and $\epsilon$ match
the results of \cite{gavrilyuk2002} derived in the context of a Baer-Nunziato-type two-fluid model.

Let us remark that in the context of bubbly flows, similar models for micro-inertia are available.
In \citep{temkin}, a pulsational energy is considered in the form: $E_{puls} = \demi M_{puls} \dot{R}^2$ and $M_{puls} = 4 \pi \rho_2 R^3$. 
An alternative approach that allows to incorporate small scale bubble velocity is also proposed in
\citep{lhuillier-nuc}, by accounting for pseudo-turbulent kinetic energies generated by
particles pulsations $K_c = \demi Q(\alpha) (D_t R)^2$, 
where $Q(\alpha)$ is proportional to $3 \alpha$ in the dilute limit of the dispersed phase. In \citep{gavrilyuk2011}, a micro-scale kinetic energy of the form $3 \alpha \rho_2 (D_t R)^2$ is used for modeling vibrations of bubbles within a bubbly flow.

Concerning $\epsilon$, this identification corresponds to the viscous contribution to damping discussed at the beginning of part \ref{par:entropy-principle}. To recover the other damping effects, that may prevail in some bubbly flow configurations, one must get rid
of the hypothesis of uniform state of the gas inside the bubbles. However, in the literature, this is only done in the linear regime
for small bubbles oscillations: this is presented in the next section.

\subsection{Connection with the linearized Rayleigh equation}
\label{section: linear R evolution}
We further examine the bubbly flow model of section~\ref{par:RP-comparison} by considering now the regime of small variations of the bubbles radius. Let us assume that 
\begin{align}
R &=\overline{R}(1 + rz)
,&
p_2 &= \overline{p_2} + r \delta p_2
,&
n &=\overline{n}
,&
\mathcal{M}_b &= \overline{\mathcal{M}_b}
,
\label{eq: RP linearization hypotheses}
\end{align}
where $\overline{R}$, $\overline{p_2}$, $\overline{\mathcal{M}_b}$, 
$\overline{n}$ are constant values and $0<r\ll 1$ is a small parameter. 
If one notes 
$
\overline{\alpha} = 4\pi \overline{R}^3 \overline{n} / 3
$, 
$
\overline{\rho_1} = \overline{n}\overline{\mathcal{M}_b} / \overline{\alpha}
$,
$
\overline{p_1} = p_1(\overline{\rho_1})
$
,
$
\overline{c_1} = c_1(\overline{\rho_1})
$
,
$
\microin(\overline{\alpha}) = \overline{\microin}
$
and
$ \overline{\rho} = \overline{\alpha} \overline{\rho_1} + (1-\overline{\alpha}) \overline{\rho_2}$
then \eqref{eq: RP linearization hypotheses} yield 
\begin{subequations}
\begin{align}
\alpha &= \overline{\alpha}(3 + rz) + O(r^2)
,&
p_1 &=\overline{p_1} - 3 \overline{\rho_1}\,\overline{c_1}^2 r z +O(r^2)
\\
\microin &= \overline{\microin} + O(r)
,&
\rho \frac{\text{d}e}{\text{d}\alpha}(\overline{\alpha}) &= 
\overline{\rho} \frac{\text{d}e}{\text{d}\alpha}(\overline{\alpha})
+3 \overline{\alpha} z 
\left[
\overline{\rho} \frac{\text{d}^2 e}{\text{d}\alpha^2}(\overline{\alpha}) 
- \overline{\rho_2} \frac{\text{d}e}{\text{d}\alpha}(\overline{\alpha}) 
\right]
r
+O(r^2)
.
\label{eq: linear RP linearized parameters}
\end{align}
\label{eq: linearized RP linearized quantities}
\end{subequations}Injecting \eqref{eq: linearized RP linearized quantities} into \eqref{eq: nonlinear ODE for R} we obtain
$$
\overline{p_1} - \overline{p_2} 
- r \delta p_2
- \overline{\rho}\frac{\text{d}e}{\text{d}\alpha}(\overline{\alpha})
-
3 \overline{\alpha} z 
\left[
\overline{\rho} \frac{\text{d}^2 e}{\text{d}\alpha^2}(\overline{\alpha}) 
- \overline{\rho_2} \frac{\text{d}e}{\text{d}\alpha}(\overline{\alpha}) 
\right]
r
= 3 r \left( 
\overline{\rho_1}\, \overline{c_1}^2 z
+\overline{\alpha}\epsilon \dot{z}
+\overline{\alpha}\,\overline{\microin}\ddot{z}
\right)
+O(r^2)
.
$$
Identifying same order terms with respect to $r$ yields
$$
\overline{p_1} - \overline{p_2} 
- \overline{\rho}\frac{\text{d}e}{\text{d}\alpha}(\overline{\alpha})
=0,
$$
and
\begin{equation}
3\overline{\alpha}\,\overline{\microin}\ddot{z}
+
3\overline{\alpha}\epsilon \dot{z}
+
3 
\left[
\overline{\rho_1}\, \overline{c_1}^2
+
\overline{\alpha}\,
\overline{\rho} \frac{\text{d}^2 e}{\text{d}\alpha^2}(\overline{\alpha}) 
- 
\overline{\alpha} \overline{\rho_2} \frac{\text{d}e}{\text{d}\alpha}(\overline{\alpha}) 
\right]
z
=
- \delta p_2 
.
\label{eq: linear RP eq}
\end{equation}
Equation \eqref{eq: linear RP eq} is a second order linear ODE in $z$ 
that is consistent with the evolution of a linear harmonic oscillator with damping and forcing terms. 
This type of equations is classic in the literature for describing the motion of vibrating bubbles in the linear regime. 
Indeed, following \cite{prosperetti77,cheng83}, we have
\begin{equation}
\overline{\rho_2} \overline{R}^2 \ddot{z}
+
2 \gamma\, \overline{\rho_2} \overline{R}^2 \dot{z}
+
\overline{\rho_2} \overline{R}^2 \omega_0^2 z
=
-\delta p_2
,
\label{eq: linear RP prosperetti - cheng}
\end{equation}
where $\gamma = \gamma_\text{vis} + \gamma_\text{th} + \gamma_\text{ac}$ drives the
damping intensity of the system. The coefficients $\gamma_\text{vis}$ pertains to viscous effects due to the surrounding liquid and is defined by $\gamma_\text{vis}= 2 \mu_2 / (\overline{\rho_2}\, \overline{R}^2)$,
$\gamma_\text{th}$ is related to 
thermal exchanges between the gas and the liquid and
$\gamma_\text{ac}$ concerns acoustic scattering by the bubbles. Expressions for $\gamma_\text{th}$ and $\gamma_\text{ac}$ are available in \cite{prosperetti77,cheng83}, and references from previous studies therein. They involve several intermediate parameters but also characteristics parameters of the forcing term, like the frequency of the perturbation $\delta p_2$. Identifying terms in \eqref{eq: linear RP eq}
and \eqref{eq: linear RP prosperetti - cheng} yields the following definitions for $\microin$ and $\epsilon$
\begin{equation}
\epsilonlin =
\frac{4\mu_2}{3\overline{\alpha}}
+\frac{2 \overline{\rho_2} \overline{R}^2}{3\overline{\alpha}}
\left(
\gamma_\text{th}
+\gamma_\text{ac}
\right)
,\quad
\mulin = \frac{\overline{\rho_2} \overline{R}^2}{3\overline{\alpha}}
.
\label{eq: def mu epsilon linear RP}
\end{equation}
We thus see that the above analysis and the resulting relation~\eqref{eq: def mu epsilon linear RP} provides a definition for $\epsilon$ that is different from \eqref{eq: evaluation of mu and epsilon}. 
More specifically, as $\epsilonlin> \epsilonRP$ we can see that \eqref{eq: def mu epsilon linear RP} yields greater damping than \eqref{eq: evaluation of mu and epsilon}. 
The discrepancy between $\epsilonlin$  and $\epsilonRP$ can be explained by simplifying hypotheses at the core of model \eqref{eq: consv system + dissipation} and our simple bubbly flow model. 
Indeed, (H3) does not allow to describe thermal exchange in the fluids and we also completely neglected pressure fluctuations within the bubbles although they importantly contribute to damping effects.

In the sequel, we will rely on the following relations:
\begin{itemize}
	\item Pfriem's expression that can be found in \cite{cheng83} for $\gamma_\text{th}$:
	\begin{equation}
	\label{eq: thermal gamma}
		\gamma_\text{th}^\text{P40} (\omega) = \omega_n \, \frac{3 (\gamma_1-1) \sqrt{2 a_1}}{2 \sqrt{\omega} R},
	\end{equation}
	where $a_1$ is the thermal diffusivity of the gas and $\gamma_1$ its ratio of specific heats. 
	\item The natural frequency $\omega_n$ is given by the relation from \cite{cheng83}:
	\begin{equation}
	\label{eq: natural frequency}
		\omega_n^2 = \frac{3 \kappa_1 p_1}{\rho_1 R^2},
	\end{equation}
	with $\kappa_1$ the thermal conductivity of gas.
	\item For $\gamma_\text{ac}$ we consider the relation found in \cite{prosperetti77}:
	\begin{equation}
	\label{eq: acoustic gamma}
		\gamma_\text{ac}^\text{P77} (\omega) = 0.5 \, \frac{\omega^2 R c_2}{c_2^2 + (\omega R)^2}.
	\end{equation}
\end{itemize}

Finally, for practical purposes, we want to get rid of the dependance of these damping parameters on the frequency of the considered 
accoustic perturbation. 
A way to get simple and constant values for these damping effects is to evaluate expressions \eqref{eq: thermal gamma} 
and \eqref{eq: acoustic gamma} at the natural frequency \eqref{eq: natural frequency} : 
\begin{equation}\label{eq:DampingCoeffAtResonance}
  \gamma_\text{th} = \gamma_\text{th}^\text{P40} (\omega_n) 
  \quad \text{ and } \quad
  \gamma_\text{ac} = \gamma_\text{ac}^\text{P77} (\omega_n).
\end{equation}

\subsection{Dispersion relations for a plane and monochromatic wave}
\label{par:dispersion relations}
	In order to test the relevance of our models and of the identification made for $\epsilon$ and $\microin$, we will compare
	in Part \ref{par:measures} the behavior of systems~\eqref{eq: consv system + dissipation}, \eqref{eq: cvv-epsilon} and \eqref{eq: CVV} in their acoustic regime to experimental measures of sound waves dispersion.

Considering smooth solutions of one-dimensional problems, all these systems can be expressed using the generic quasilinear form 
\begin{equation}
\partial_t \mathbf{W} 
+ \mathbf{A}(\mathbf{W})\partial_x \mathbf{W} = \mathbf{S}(\mathbf{W})
.
\label{eq: general conslaw w source}
\end{equation}
Following standard lines \citep{whitham, burman}, we seek for a monochromatic wave solution of \eqref{eq: general conslaw w source} by writting $\mathbf{W}$ in the form
\begin{equation}
\mathbf{W}(x,t) = 
\mathbf{W}^{(0)}
+
r
\mathbf{W}^{(1)}(x,t)
+
O(r^2),
\quad
\mathbf{W}^{(1)}(x,t)
=
\widehat{\mathbf{W}}^{(1)}
\exp \Big( i\omega t - i k(\omega)x \Big)
,
\label{eq: monochromatic wave form}
\end{equation}
where $\omega$ is the angular frequency, $k$ the wavelength and $r$ is a small amplitude parameter. 
The states $\widehat{\W}^{(1)}$ and $\W^{(0)}$ are both constant. The fluid parameters involved with $\W^{(0)}$ are noted with the superscript ${}^{(0)}$ and for the sake of simplicity, we suppose that $\W^{(0)}$ is always a rest state, \text{i.e.} $u^{(0)} = 0$.

Injecting \eqref{eq: monochromatic wave form} into \eqref{eq: general conslaw w source} and 
identifying terms with respect to the powers of $r$
yields
\begin{equation}
\mathbf{S}(\mathbf{W}^{(0)}) = 0
,\quad
\widehat{\mathbf{W}}^{(1)}
\in
\ker\Big(
i\omega \mathbf{Id} 
- ik(\omega)\mathbf{A}(\mathbf{W}^{(0)}) 
- \mathbf{S}'(\mathbf{W}^{(0)}) 
\Big).
\end{equation}
Consequently $\omega$ and $k(\omega)$ are bound by the so-called dispersion relation
\begin{equation}
\det\Big(
i\omega \mathbf{Id} 
- ik(\omega)\mathbf{A}(\mathbf{W}^{(0)}) 
- \mathbf{S}'(\mathbf{W}^{(0)}) 
\Big) = 0,
\label{eq: general dispersion relation}
\end{equation}
which allows to defined the phase velocity and the spatial attenuation of the acoustic wave respectively by $\mathfrak{Re}[\omega / k(\omega)]$ and 
$\mathfrak{Im} [k(\omega)]$.
Let us now detail the results for each system of our hierarchy.
Let us note
 $$
 \ds{
 \mathsf{H} (\rho, Y, \alpha) 
 = 
 \rho  
 \frac{Y(1-Y) c_1^2 (\rho_1) c_2^2(\rho_2)}{\alpha^2 (1-\alpha)^2}}
 .
 $$
\begin{itemize}
\item
For the two-phase model with micro-inertia~\eqref{eq: consv system + dissipation}, 
the dispersion relation, the associated phase velocity $\cPhaseMuEpsilon$ and the spatial attenuation $\betaMuEpsilon$ read:
\begin{align}
\left(
	\frac{\kMuEpsilon(\omega)}{\omega} 
\right)^2
	&= 
	\frac{
	\microin   \omega^2   -    i  \epsilon   \omega   -  \,\cWood^{-2} 
	 \mathsf{H} (\rho^{(0)},Y^{(0)},\alpha^{(0)})
	}
	{
	\microin  \cFrozen^2   \omega^2   -    i  \epsilon \,\cFrozen^2   \omega   -  \, 
	\mathsf{H} (\rho^{(0)},Y^{(0)},\alpha^{(0)})
	},
&
\cPhaseMuEpsilon(\omega) 
&= 
\mathfrak{Re} \left[\frac{\omega}{\kMuEpsilon(\omega)}\right]
,
&
\betaMuEpsilon(\omega) 
&=
\mathfrak{Im}[\kMuEpsilon(\omega)]
.
\label{eq:CVV-mu-epsilon-dispers}
\end{align}
\item
For the micro-inertia free model~\eqref{eq: cvv-epsilon} obtained by the limit $\microin\to 0$, $\epsilon=O(1)$, we get the dispersion relation, phase velocity $\cPhaseEpsilon$ and attenuation $\betaEpsilon$ defined by
\begin{align}
	\left(
	\frac{\kEpsilon(\omega)}{\omega}
	\right)^2
	& = 
	\frac{
	i \epsilon   \omega +  \cWood^{-2}  \mathsf{H} (\rho^{(0)},Y^{(0)},\alpha^{(0)}) 
	}{
	i \epsilon   \cFrozen^2   \omega +
	\mathsf{H} (\rho^{(0)},Y^{(0)},\alpha^{(0)})
	}
,&
\cPhaseEpsilon(\omega) 
=&
 \mathfrak{Re} \left[\frac{\omega}{\kEpsilon(\omega)}\right]
,&
\betaEpsilon(\omega) 
=&
 \mathfrak{Im}[\kEpsilon(\omega)]
 .
\label{eq:CVV-epsilon-dispers}
\end{align}
\item
Finally for the full-equilibrium model~\eqref{eq: CVV} when $\microin\to 0$, $\epsilon=\petitO(\sqrt{\microin})$, the dispersion relation reads
\begin{align} 
\label{eq:CVV-dispers}
	\frac{k(\omega)^2}{\omega^2} &= \frac{1}{ \cWood^2}
	,&
	\cPhase(\omega) &= \cWood
	,&
	\beta &= 0.
\end{align}
\end{itemize}
We recall that $\cWood$ and $\cFrozen$ are defined by \eqref{eq:cWood} and \eqref{eq:cFrozen}.

We observe that when one accounts for internal damping with $\varepsilon>0$  and with micro-inertia (resp. without micro-inertia), the phase velocity of the acoustic wave $\cPhaseMuEpsilon$ (resp. $\cPhaseEpsilon$)
is not equal to the sound velocity $c$ (resp. $\cFrozen$) issued from the characteristic velocities when one discards the source terms in the system. This underlines the fact that micro-inertia and damping source terms have a substantial influence on the phase velocity of the acoustic wave.

\begin{figure}
	\begin{center}
		\subfloat[\small Phase velocity.]{
			\begin{minipage}[c]{
				0.5\textwidth}
				\includegraphics[width=8.5cm]{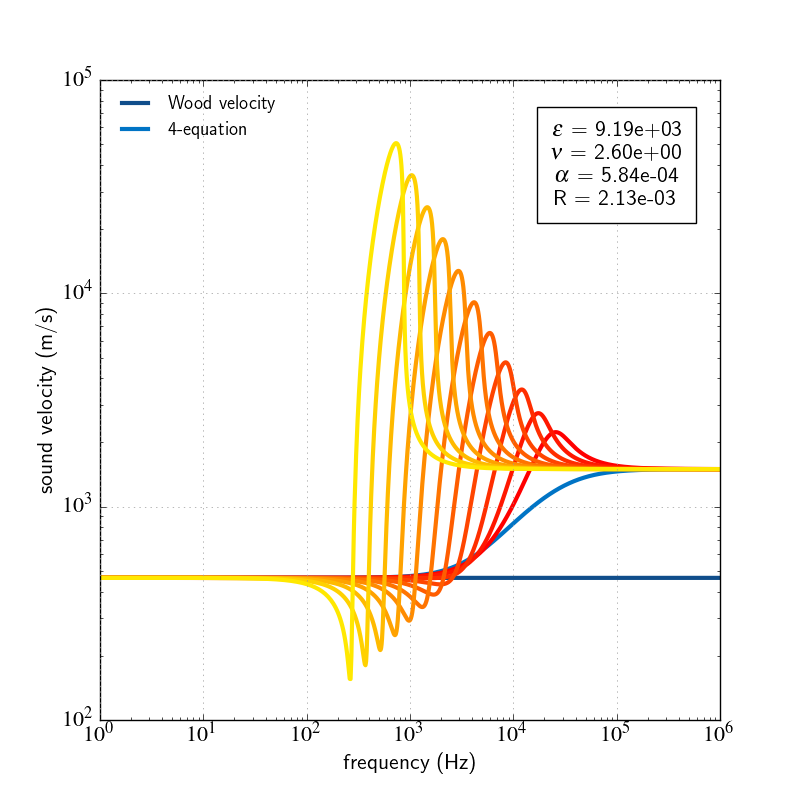}
			\end{minipage}}
		\subfloat[\small Spatial attenuation.]{
			\begin{minipage}[c]{
				0.5\textwidth}\includegraphics[width=8.5cm]{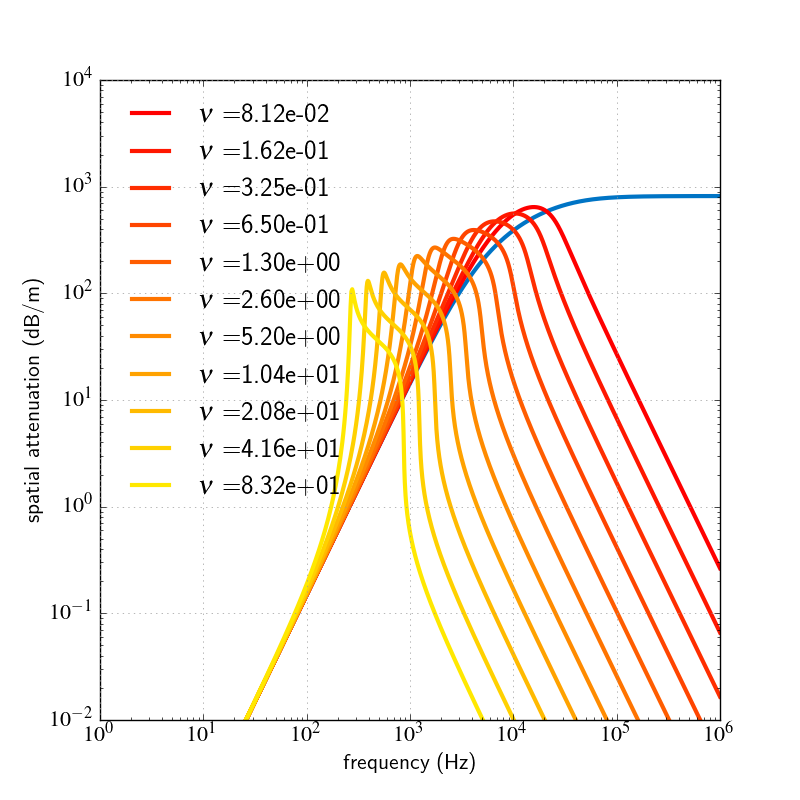}
			\end{minipage}}
		\caption{Phase velocity and spatial attenuation for the 5-equation model and influence of the value of $\microin$.}
		\label{fig:influence-mu}
	\end{center}
\end{figure}

\begin{figure}
	\begin{center}
		\subfloat[\small Phase velocity.]{
			\begin{minipage}[c]{
				0.5\textwidth}
				\includegraphics[width=8.5cm]{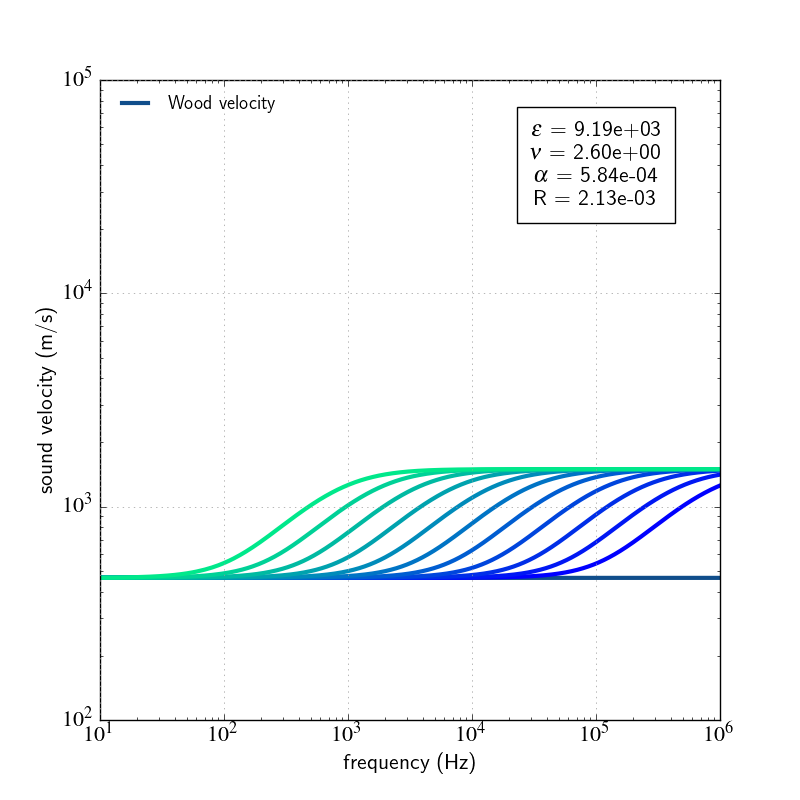}
			\end{minipage}}
		\subfloat[\small Spatial attenuation.]{
			\begin{minipage}[c]{
				0.5\textwidth}\includegraphics[width=8.5cm]{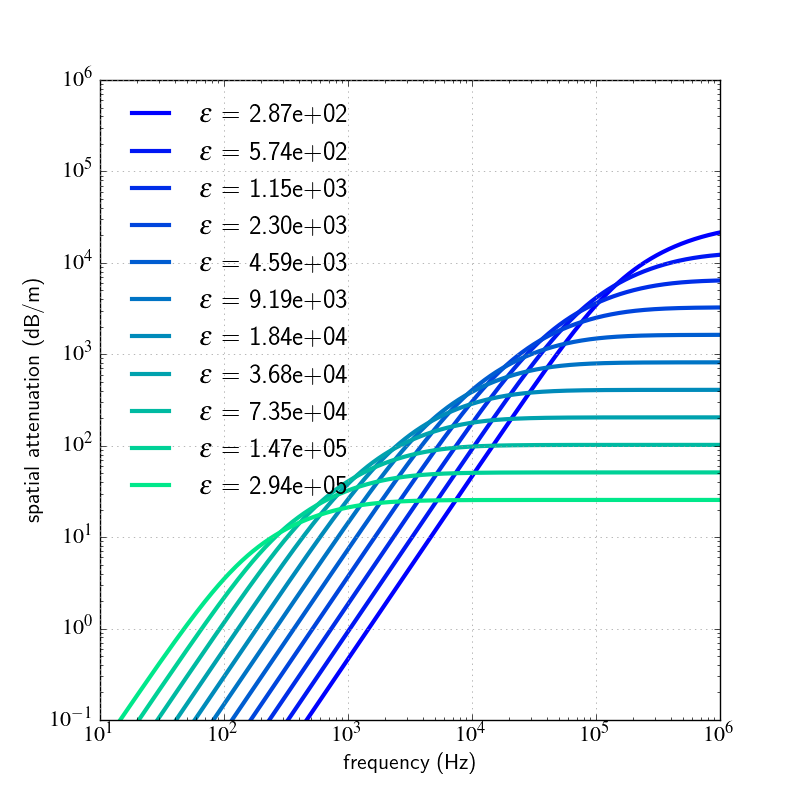}
			\end{minipage}}
		\caption{Phase velocity and Spatial attenuation for the 4-equation model and influence of the value of
		 $\epsilon$.}
		\label{fig:influence-epsilon}
	\end{center}
\end{figure}

Let us now examine the variations of the dispersion relations across the hierarchy.  As $\microin$  and $\varepsilon$ reach their asymptotic limit, the
transition from one model to another materializes through the dispersion relations~\eqref{eq:CVV-mu-epsilon-dispers}, \eqref{eq:CVV-epsilon-dispers} and \eqref{eq:CVV-dispers}.
Indeed, we see that
\begin{align*}
\lim_{\microin\to 0}
\cPhaseMuEpsilon
&=
\cPhaseEpsilon
,&
\lim_{
\substack{\microin\to 0 \\ \varepsilon = o(\sqrt{\microin})}
}
\cPhaseMuEpsilon
&=
\cPhase
,&
\lim_{\varepsilon \to 0}
\cPhaseEpsilon
&=
\cPhase
.
\end{align*}
Le us note that these limits are not uniform over all frequencies: 
this is illustrated in figures \ref{fig:influence-epsilon}. Indeed, we can observe that, whatever the value of $\varepsilon$,
there is a frequency above which $\cPhaseEpsilon$ will be close to $\cFrozen$. However, this critical frequency increases
when $\varepsilon$ decreases. On the contrary, the transition from $\cPhaseMuEpsilon$ to $\cPhaseEpsilon$ is more uniform,
as illustrated in figures \ref{fig:influence-mu}. 
Indeed, when $\microin$ decreases, the damping effects due to $\varepsilon$ prevail, and although  the resonant frequency
only depends on $\microin$, the effects of resonance are completely attenuated because of $\varepsilon$. One can also
note that the asymptotic behavior for spatial attenuation at low frequencies does not depend on the value of $\microin$, but
depends on the value of $\varepsilon$, see figure \ref{fig:influence-epsilon} for the 4-equation model.

The model hierarchy also shows through when one spans frequency values $\omega$ (and can also be noticed on Figures 
\ref{fig:influence-mu} and \ref{fig:influence-epsilon}). 
Indeed, if one considers acoustic waves at low frequencies $\omega\ll 1$, then the dispersion relations~\eqref{eq:CVV-mu-epsilon-dispers},
\eqref{eq:CVV-epsilon-dispers} and \eqref{eq:CVV-dispers} yield
\begin{align*}
\left(
\frac{\kMuEpsilon(\omega)}{\omega}
\right)^2
&=
\left(
\frac{\kEpsilon(\omega)}{\omega}
\right)^2
+O(\omega^2)
=
\frac{1}{\cWood^2} + O(\omega)
\end{align*}
In terms of phase velocities and attentuation we obtain
\begin{align*}
\cPhaseMuEpsilon
&=
\cPhaseEpsilon
+
O(\omega^2)
=
\cWood + O(\omega)
,&
\betaMuEpsilon
=
\betaEpsilon
+O(\omega^4)
=O(\omega)
.
\end{align*}
In the limit $\omega\to 0$, the phase velocity of the acoustic waves for all models will tend to $\cPhase=\cWood$ 
and the spatial attenuation will vanish.
Let us now turn to high frequencies $\omega \gg 1$. From  \eqref{eq:CVV-mu-epsilon-dispers},
\eqref{eq:CVV-epsilon-dispers} and \eqref{eq:CVV-dispers} we have
\begin{equation}
\left(
\frac{\kMuEpsilon(\omega)}{\omega}
\right)^2
=
\left(
\frac{\kEpsilon(\omega)}{\omega}
\right)^2
+O\left(\frac{1}{\omega^2}\right)
=
\frac{1}{\cFrozen^2}
+O\left(\frac{1}{\omega}\right)
.
\end{equation}
Thus the acoustic waves of both systems equipped with internal damping are such that
\begin{equation}
\cPhaseMuEpsilon  
=
\cPhaseEpsilon + O\left(\frac{1}{\omega^2}\right)
=
\cFrozen
+O\left(\frac{1}{\omega}\right)
.
\end{equation}
Consequently, in the limit $\omega\to + \infty$, 
the phase velocities of the acoustic waves associated with 
\eqref{eq: consv system + dissipation} and \eqref{eq: general consv system}
tend to $\cFrozen$, while it remains constant and equal to $\cWood$ for \eqref{eq: CVV}. In all cases the spatial attenuation tends to $0$.

Finally, let us mention a distinctive behavior of the acoustic waves for the system~\eqref{eq: consv system + dissipation} equipped with both damping and micro-inertia : in the case of low internal dissipation \textit{i.e.} small values of $\varepsilon$, the dispersion relation~\eqref{eq:CVV-mu-epsilon-dispers}  leads to resonance in the vicinity of the frequency 
\begin{equation}
\label{eq:Resonance}
\resofreq = 
\inv{\cWood} 
\sqrt{
\frac{ \mathsf{H} (\rho^{(0)},Y^{(0)},\alpha^{(0)} )} {\microin} 
}
.
\end{equation}

\section{Comparison with bubbly flow reference data }
\label{par:measures}

From now on, we shall suppose that the compaction energy is null $e(\alpha) = 0$, 
and that the barotropic EOS for each pure fluid have the form
\begin{equation}
p_k = p_k^0 + c_k^2 \, (\rho_k - \rho_k^0),
\label{eq: pure fluid EOS}
\end{equation}
where $c_k$, $\rho^0_k$ and $p_k^0$ are real constants chosen as follows
\begin{align*}
p_1^0 &= 1.0 \times 10^5 \,\text{Pa}
,&
c_1 &= 340 \,\text{m}.\text{s}^{-1}
,&
\rho^0_1 &= 1.2 \,\text{kg}.\text{m}^{-3},
\\
p_2^0 &= 1.0 \times 10^5 \,\text{Pa}
,&
c_2 &= 1500 \,\text{m}.\text{s}^{-1}
,&
\rho^0_2 &= 1000 \,\text{kg}.\text{m}^{-3}
.
\end{align*}

Thanks to the dispersion relations \eqref{eq:CVV-mu-epsilon-dispers}, 
\eqref{eq:CVV-epsilon-dispers} and \eqref{eq:CVV-dispers}, we can now
study the response of 
systems~\eqref{eq: consv system + dissipation}, 
\eqref{eq: cvv-epsilon} and \eqref{eq: CVV} in the acoustic regime under a forced pressure oscillation and compare them with experimental results obtained for bubbly flows. The data we shall use rely on two experimental works by \citep{silberman} and \citep{leroy08}. Let us first briefly outline the framework of these studies. In the sequel, elements related to the experimental data will be denoted with the superscrit ${}^\text{ref}$.

The data of \citep{silberman} are considered as reference experimental results in the domain of acoustic wave propagation for bubbly flows. They have been used in comparisons with several models in the literature \citep{wijngaarden, cheng85, commander, drew-passman}.
The method proposed in \citep{silberman} consists in generating standing waves in various length steel pipes. The sound is generated at one end of the pipe while small hydrophones measure sound pressure at the other end. Measures are performed between two nodes or antinodes that allow to compute the phase velocity and the spatial attenuation.
The size distribution of the bubbles is estimated using photographs.
The resulting measures were very accurate, except near the resonance frequency $\resofreqref$. Indeed, for $\omega$ close to $\resofreqref$ 
evaluation of the phase velocity was not possible due to the severe attenuation of the acoustic waves. In order to obtain data in this range of frequencies, we shall use the work of \citep{leroy08} that involves acoustic wave propagating within a thin hair gel sample  containing air bubbles. The sound waves are produced at one end of the system by a transducer and measurements are performed thanks to an hydrophone at the other end. 
The advantage of using the gel is that the distribution of bubbles radii and volume fractions are accurately known.
According to \citep{leroy08}, the difference in terms of acoustic behavior between water and gel is negligible regarding the wave dispersion. 
Thanks to this set up, the results of \citep{leroy08} provide accurate data for both  phase velocity and attenuation in the vicinity of $\resofreqref$.

\subsection{Influence of micro-viscosity and micro-inertia in the acoustic regime}
\label{section: data comparison 1}

We shall examine the behavior of the models when $\omega$ spans the possible frequencies and distinguish three main ranges of frequencies for characterizing the phase velocity and the spatial attenuation. Then we will focus specifically on the near-resonance frequencies.

\subsubsection*{Comparison across the whole spectrum of frequencies $\omega$}
\begin{figure}
	\begin{center}
		\subfloat[\small Phase velocity.]{
			\begin{minipage}[c]{
				0.5\textwidth}
				\includegraphics[width=8.5cm]{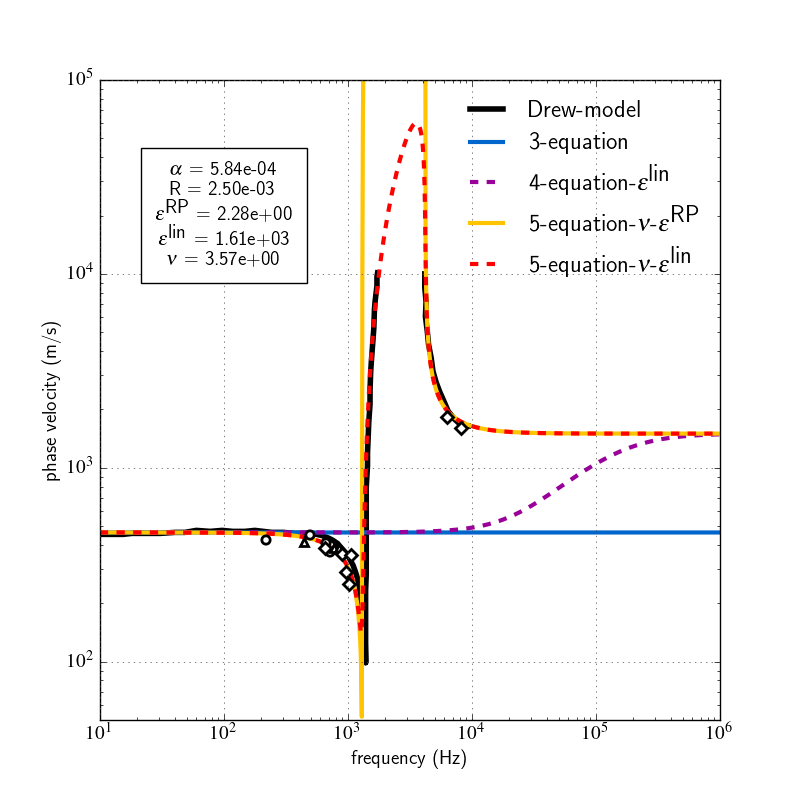}
			\end{minipage}}
		\subfloat[\small Spatial attenuation.]{
			\begin{minipage}[c]{
				0.5\textwidth}\includegraphics[width=8.5cm]{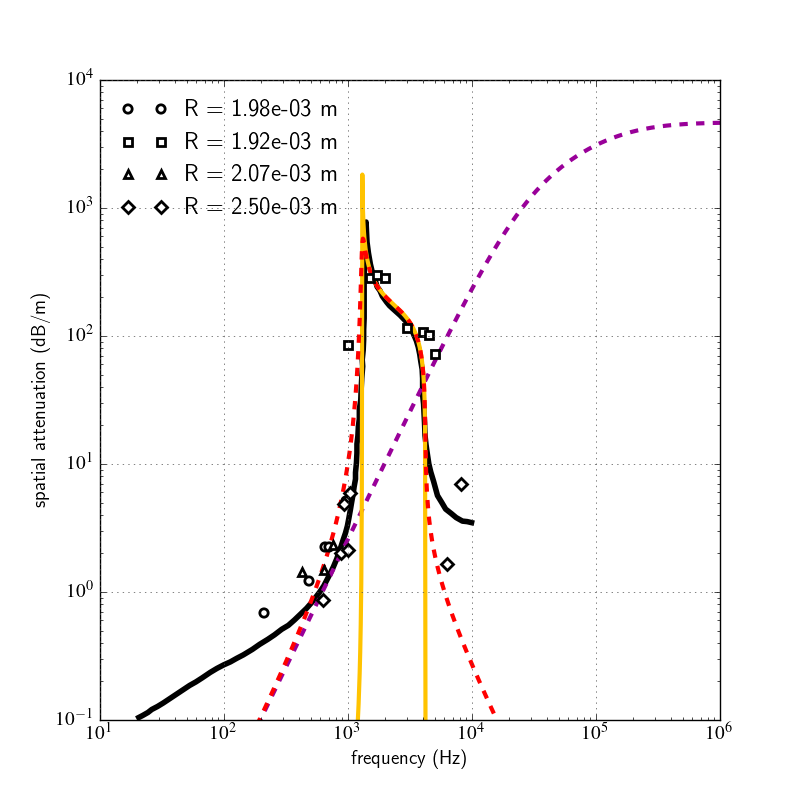}
			\end{minipage}}
		\caption{Dispersion relations for the different models from the hierarchy (full and dashed lines)
		and Silberman's measures (symbols) for radii of bubbles around $R = 2.0 \times 10^{-3}$ m, $\alpha = 5.84\times10^{-4}$.}
		\label{fig:silberman-small1}
	\end{center}
\end{figure}

We consider a set of measures from \citep{silberman} that involve a flow  characterized by  $R=2.5\,\text{mm}$ and $\alpha = 5.84 \times 10^{-4}$.
Using relations \eqref{eq: evaluation of mu and epsilon} and \eqref{eq: def mu epsilon linear RP}, we obtain  the following values for the parameters of the 5-equation model~\eqref{eq: consv system + dissipation} and the 4-equation model~\eqref{eq: cvv-epsilon}
\begin{align}
\epsilonRP &=   2.28 
, & 
\muRP = 3.57,
\label{eq: mu epsilon estimate RP }
\\
\epsilonlin &=   1.61\times 10^3 
, & 
\mulin = 3.57.
\label{eq: mu epsilon estimate lin}
\end{align}
Figure~\ref{fig:silberman-small1} displays both phase velocity and spatial attenuation for all models of the hierarchy, for the model of Drew~\citep{cheng85}, superposed on the experimental results.

\begin{itemize}
	\item \textbf{Range $\omega \ll \resofreqref$.}
	For low frequencies, the pressure perturbation is very slow and thus we can expect the bubbles of the system to remain at an equilibrium state with respect to both mechanics and thermodynamics. Little internal dissipation is involved with this regime, which is visible through the measures that show a low spatial attenuation. The evaluation of $\cPhaseref$ in the experiment provides values that are close to $\cWood$. All the models of the hierarchy show a good agreement with these data, as seen in Fig.~\ref{fig:silberman-small1}.
	As $\omega$ increases, the spatial attenuation $\betaref$ also increases. This trends is correctly followed by the models of the hierarchy that account for internal dissipation. Nevertheless, \cite{prosperetti77} underlines that in this regime and up to a certain frequency, the thermal dissipation is the dominant internal dissipation effect. 
	For all models, $\betaMuEpsilon$ and $\betaEpsilon$ are lower than experimental data. 
  However, if the five-equation model~\eqref{eq: consv system + dissipation} importantly underestimates $\betaMuEpsilon$ 
  when the damping is matched on Rayleigh-Plesset $(\epsilon=\epsilonRP)$, 
  the match with the reference data is very good when $\epsilon=\epsilonlin$.
	\item \textbf{Range $\omega$ close to $\resofreqref$.} Near resonance, $\betaref$ increases with $\omega$ and becomes very large. On the contrary, $\cPhaseref$ decreases as $\omega$ increases. It reaches $\cPhaseref = 0$ for some frequency $\extinctfreqref < \resofreqref$. For $\omega\in[\extinctfreqref,\resofreqref]$, there is a good agreement between
	$\cPhaseMuEpsilon$, $\cPhaseEpsilon$ and $\cPhaseref$. 
		On the contrary, there is an important discrepancy between the phase velocity predicted by the model of Drew and $\cPhaseMuEpsilon$, $\cPhaseEpsilon$ for this range of frequencies. It is worth noting that 
	$\cPhaseMuEpsilon$ is much closer to the reference value for $\epsilon=\epsilonlin$ than for $\epsilon=\epsilonRP$.
		Concerning the spatial attenuation, we can see that models of the hierarchy that account for micro-inertia, namely \eqref{eq: consv system + dissipation} and \eqref{eq: general consv system} fit quite well the reference results. 
  In contrast, models \eqref{eq: cvv-epsilon} and \eqref{eq: CVV} yield a poor estimate of the spatial attenuation.
	This suggests that the source terms related to $\microin$ in \eqref{eq: general system +dissip eq alpha} and \eqref{eq: general system + dissip eq w} play a key role in the system behavior when $\omega \approx \resofreqref$. 
	Moreover it also hints that our estimate for $\microin$ in \eqref{eq: evaluation of mu and epsilon} and \eqref{eq: mu epsilon estimate lin} is coherent. Finally, the results suggest that the source terms related to $\epsilon$ in \eqref{eq: general consv system} do not have a great influence on the values of $\betaMuEpsilon$ in this range of frequencies.	
	\item \textbf{Range $\omega \gg \resofreqref$.} For high frequencies, very few experimental data are available and thus the main comparison elements are given by the model of Drew. In this regime, one can presume that acoustic radiation effects cannot be neglected. Indeed, the bubbles start emitting acoustic waves that are transmitted to the liquid. This process will remove energy from the bubbles and therefore will become the main damping effect of the system. For all systems of our hierarchy except \eqref{eq: CVV}, the phase velocity will tend to $\cFrozen$, which agrees with the behavior of Drew's model. The spatial attenuation coefficients
	of the model hierarchy do not match well the reference data of Drew's model in this range of frequencies: 
  \eqref{eq: cvv-epsilon} clearly overestimates damping (purple dashed line), 
  when model \eqref{eq: consv system + dissipation} with matching coeffictients $\muRP$ and $\epsilonRP$ provides too low a dissipation 
  (yellow plain line).
  For the same model, the micro-viscosity choice $\epsilonlin$ clearly increases the damping effect 
  but still at a much lower level than the dissipation of Drew's model (red dashed line).
\end{itemize}

\subsubsection*{Finer comparison near resonance}

In Leroy's experiment \citep{leroy08}, the set of measures is very dense for $\omega$ close to $\resofreqref$. In this paragraph we discard both 4-equation model~\eqref{eq: cvv-epsilon} and 3-equation model~\eqref{eq: CVV} as they cannot produce resonant behavior.
The bubbles in \citep{leroy08} are smaller than those of \citep{silberman}, we thus consider different values of $(R,\alpha)$ by setting  $R \approx 8.1 \times 10^{-5}\,\text{m}$ and $\alpha = 1.5 \times 10^{-4}$. Thanks to  \eqref{eq: evaluation of mu and epsilon} and \eqref{eq: def mu epsilon linear RP} we obtain
\begin{align}
\epsilonRP &=  8.89\times 10^{-1} 
,&
\muRP &=1.46\times 10^{-3}.
\label{eq: mu epsilon estimate leroy RP}
\\
\epsilonlin &=  87.3 
,&
\mulin &=1.46\times 10^{-3}.
\label{eq: mu epsilon estimate leroy lin}
\end{align}

\begin{figure}
	\begin{center}
		\subfloat[\small Phase velocity.]{
			\begin{minipage}[c]{
				0.5\textwidth}
				\includegraphics[width=8.5cm]{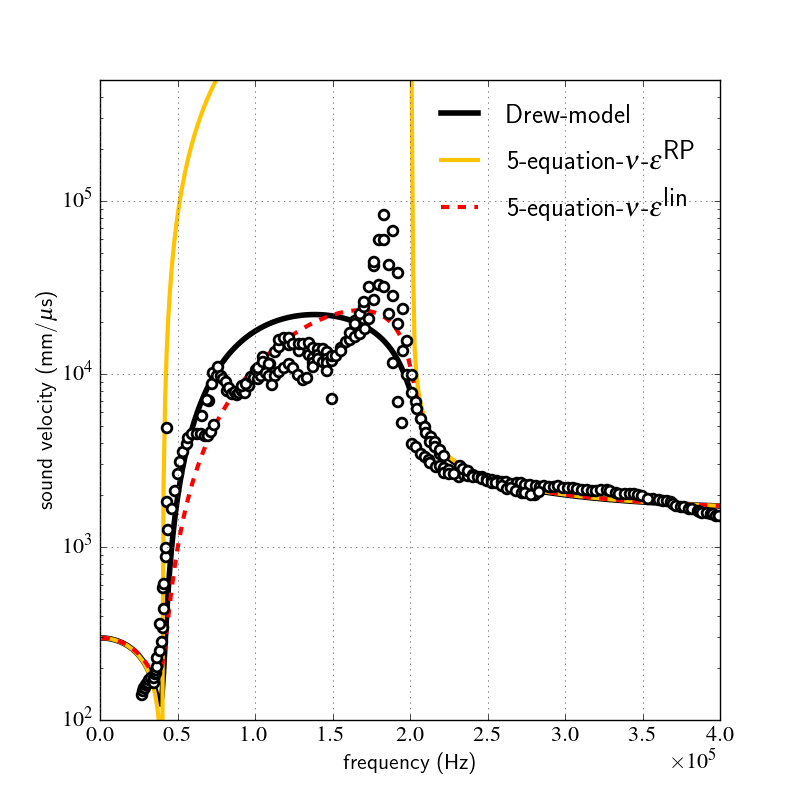}
			\end{minipage}}
		\subfloat[\small Spatial attenuation.]{
			\begin{minipage}[c]{
				0.5\textwidth}\includegraphics[width=8.5cm]{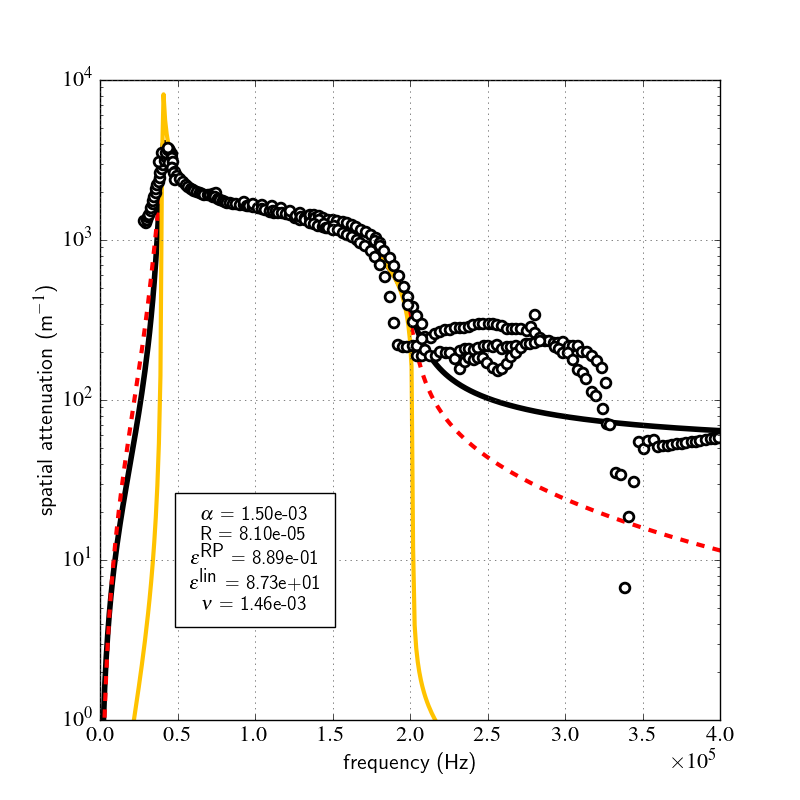}
			\end{minipage}}
		\caption{Dispersion relations for the different models from the hierarchy (full and dashed lines)
		and Leroy's measures (symbols) for radii of bubbles around $R = 8.1 \times 10^{-5}$ m, $\alpha = 1.5\times10^{-4}$.}
		\label{fig:leroy-small1}
	\end{center}
\end{figure}

The results we obtain with this set of parameters is coherent with the previous 
comparison. Indeed, in figure \ref{fig:leroy-small1} we can see that
for $\omega$ close to $\resofreqref$ the phase velocity $\cPhaseMuEpsilon$  
is clearly overestimated for $\epsilonRP$ but the match with experimental data for $\epsilonlin$ seems more accurate than the for model of Drew. Regarding the attenuation $\betaMuEpsilon$, the choice of $\epsilonlin$ gives clearly a better match with the reference data than $\epsilonRP$.

\subsection{On the evaluation of the micro-viscosity and the micro-inertia}
We succeeded in identifying the micro-inertia $\microin = \mulin = \muRP$ and the micro-viscosity $\epsilon=\epsilonRP$ (resp. $\epsilon=\epsilonlin$) in the five-equation model~\eqref{eq: consv system + dissipation} thanks to comparison with the Rayleigh-Plesset equation~\eqref{eq: nonlinear ODE for R} (resp. the linear radius evolution equation~\eqref{eq: linear RP eq}). However, the values of $\epsilonRP$ and $\epsilonlin$ are significantly different and for a given $\omega \geq \resofreqref$, $\epsilonRP$ and $\epsilonlin$ yield different acoustic behaviors for system~\eqref{eq: consv system + dissipation}: the agreement with reference data is better for $\epsilon=\epsilonlin$. We believe that underlying hypotheses of our model hierarchy clearly impact the definition of $\epsilon$.
Indeed, assumption (H6) does not allow to account for pressure fluctuations within the bubble, neither for pressure fluctuations in the liquid that act as inertial terms with respect to the bubble motion. These phenomena are the ground for thermal and acoustic damping effects, that are respectively driven by $\gamma_\text{ac}$ and $\gamma_\text{th}$, see \cite{prosperetti77} and \cite{cheng83}. 

What we observed in section~\ref{section: data comparison 1} is coherent with the analysis of \cite{prosperetti77}: for the flows settings we chose, both thermal and acoustic damping are the main damping effects involved in the bubble vibrations. According to \cite{prosperetti77}, when the radius is close to the value of the Silberman experiment, $R=1\,\text{mm}$, one can estimate that $\gamma_\text{th}/\gamma_\text{vis} \in[5,5000]$ for $\omega \leq 10^5\,\text{s}^{-1}$ and $\gamma_\text{ac}/\gamma_\text{vis} \in[5,5000]$ for $\omega \geq 5	\times 10^3\,\text{s}^{-1}$. When 
$R=8.1\times 10^{-2}\,\text{mm}$, a close value to the bubbles radii in Leroy's experiment, we have that $\gamma_\text{th}/\gamma_\text{vis} \in[10,500]$ for $\omega \leq 10^6 \,\text{s}^{-1}$ and $\gamma_\text{ac}/\gamma_\text{vis} \in[3,500]$ for for $\omega \geq 10^5 \,\text{s}^{-1}$.

Nevertheless, we saw that, by increasing the value of $\betaMuEpsilon$, it is possible to enforce an equivalent damping $\betaMuEpsilon \approx \beta_\text{th}$ that better fits the reference data in this range of frequencies, especially close to resonance. 
This enables an alternate mean for determining $\epsilon$ and $\microin$ in order to cope with the flaws of the model without revisiting the core assumptions of the model.  One could tune $(\epsilon,\microin)$ in a heuristic way to better fit $\cPhaseMuEpsilon$ and $\betaMuEpsilon$ with respect to reference data, and thus use the acoustic regime behavior of model~\eqref{eq: consv system + dissipation} as an evaluation tool.

\section{Conclusion}

In this work, we have proposed a hierarchy of 3 compressible two-phase flows models.
We have proceeded by deriving a model for a compressible barotropic two-phase medium that accounts for both bulk and small-scale vibrational kinematic phenomena. A notable characteristic of this model is that it is agnostic with respect to the regime in the sense that no assumption is made \textit{a priori} concerning the topology of the interfaces: 
the model may describe either separate-phase or disperse-phase flows. 
Following classical modeling guidelines, we used the Least Action Principle 
and the elaboration of an entropy budget, to obtain respectively the conservative and dissipative structures of our model.

The resulting system is a five-equation model whose convective part is hyperbolic. 
This system features two parameters: $\epsilon$ that is related to internal dissipation effects 
and $\microin$ that pertains to small-scale kinematic effects. 
Two reduced models can be obtained by considering the regimes $\microin\to 0$, $\epsilon=O(1)$ and $\microin\to 0$, $\epsilon=O(\sqrt{\microin})$. 
These limit regimes led to models from the literature that have been used for describing compressible separate-phase flows~\citep{ONERA-chante,champmartin2014}. 
Then we equipped our five-equation model with additional flow structure hypotheses that enabled the description of monodisperse bubbly flows. 
This allowed us to recover two evolution equations for the bubble radius in the nonlinear and linear regime, which can be interpreted as a complete and a linearized Rayleigh-Plesset equation. 
These connections have brought two important elements: first it has suggested to relate $\epsilon$ to micro-viscosity effects and $\microin$ to micro-inertial ones. 
Then, it has also allowed to propose two possible evaluations for $(\epsilon, \microin)$: a first evaluation that is based on the nonlinear Rayleigh-Plesset equation and a second one that uses the linearized Rayleigh-Plesset equation. Comparing both evaluations allowed to shed some light on the impact of the 
simplifying hypotheses used for modeling the behavior of the fluid within the bubbles. The second evaluation suggested a way to compensate for neglected effects that may even be predominant in some flow configurations.

We finally considered flows in the acoustic regime: we compared the behavior of monochromatic acoustic waves for each models of the regime with reference data provided by both experimental data and the two-phase model of \citep{drew-passman}.
First, this study allowed to further discriminate the domain of validity of each model of the hierarchy. Moreover, this comparison highlighted differences between the models of the hierarchy like the ability of obtaining resonance regime. Good matches with the reference data were obtained with the five-equation model and validated the evaluations of $(\epsilon, \microin)$ in this particular situation. In this way, this study brought to light that the acoustic regime can be used as a useful modeling tool for testing and tuning key elements in flow models.

The present work may be extended by accounting for additional phenomena like thermal effects. The model hierarchy that was proposed here may also be exploited for investigating coupling strategies  in the context of numerical simulations involving different regimes.

\begin{appendices}

\makeatletter{}\section{The Drew-Passman Bubbly Flow Model}
We recall hereafter the two-phase bubbly system studied in \citep{drew,cheng85}.  For the sake of simplicity, we shall only consider one-dimensional problems.
This model is compound of balance equations for mass, momentum and energy for each fluid.
These equations are derived following the lines of  \citep{ishii} for deriving flow-field equations using averaging methods.
Constitutive equations are also used for closure purposes, and finally, an additional interaction laws enables to 
relate the pressures of each component and close definitely the equations. 
The derivation of this interaction law is presented just below, and is based on the same approach that the one
used for deriving Rayleigh-Plesset's equation \citep{rayleigh,plesset77}.

First, consider the motion of a single bubble with a spherical shape of radius $R$:
one makes the assumption that the bubble undergoes oscillations that are driven by a velocity potential $\phi$ of the form
$$ 	
	\varphi(r) = - \frac{R^2 \dot{R}} {r(1-ikR)}\exp(ik(r-R)),
$$
where $r$ is the distance to the center of the bubble (see \citep{landau}) and $k$ is the wave number associated with the disturbance of the velocity field coming from the surrounding liquid. 
The evolution of $R$ is derived by supposing that, in comparison with the gas inside the bubble, the liquid is almost incompressible and that the pressure far from the bubble interface is $p_2$, then one supposes the dynamics of the bubble to verify the Bernoulli equation as follows
\begin{equation}
 	\frac{p_{2i}(t)}{\rho_2} \, + \, \demi \left( \grad \varphi(R) \right) ^2 \, + \, \partial_t \varphi(R)  \, = \, \frac{p_2(t)}{\rho_2}
 	,
 	\label{eq: bernoulli one bubble}
\end{equation}
where $p_{2i}$ and $\rho_2$ are respectively the pressure of the liquid at the interface between fluids and the density of the liquid.
Moreover, one supposes that the motion of the bubble is constrained by the Laplace relation that gives the jump relation between the pressure of both fluids across the interface of the bubble
\begin{equation}
	p_{1i} - p_{2i} = \frac{2 \sigma}{R} - 2 \mu_2 \partial_r u_2 |_{r=R}
 	\label{eq: Laplace jump relation}
\end{equation}
where $\sigma$ is the surface tension and $\mu_2$ is the dynamic viscosity of the liquid and $p_{1i}$ is the pressure of the gas at the bubble boundary.
Supposing the amplitude of the bubble oscillations to be small, \eqref{eq: Laplace jump relation} and \eqref{eq: bernoulli one bubble} are complemented by an additional relation that connects the gas interfacial pressure variations $\delta p_{1i}$ to the radius variations $\delta R$.
This relation accounts for thermal effects occurring at the interface and also for the thermodynamics properties of the gas. It involves complex expressions that will not be detailed here, we refer the reader to \citep{prosperetti77, cheng83} for a detailed view on this topic.

The other part of Drew's model is made of six bulk balance equations given below 
\eqref{eq: DP system m1}-\eqref{eq: DP system m2h2}.
They characterize  the evolution of the system parameters at the macroscopic scale and are derived by applying an \textit{ensemble averaging}~\citep{drew-passman}. 
Let us note $\rho_q$, $u_q$, $p_q$, $h_q$ respectively the 
averaged values of
density, velocity, partial pressure and specific enthalpy of the fluid $q=1,2$. Each fluid is supposed to be a compressible material that is equipped with a pressure law of the form $(\rho_k, h_k)\mapsto p_k$.
 In the case of a dispersed bubbly flow, the volume fraction of gas is defined by setting $\alpha = 4 \pi n R^3 / 3$, where $n$ is the bubble number density. The partial masses are given by
 $m_q = \rho_q \alpha_q$, where $\alpha_1 = \alpha$, $\alpha_2 = 1-\alpha$.
  Neglecting wall-shear effects and gravity, for one-dimensional problems the system reads as follows
\begin{subequations}
\begin{align} 
		\partial_t \, m_1  +  
		\partial_x \, (m_1 u_1)  = & 0 ,
		\label{eq: DP system m1}
		\\
		\partial_t \, m_2  +   \partial_x \, (m_2 u_2)  = & 0 ,
		\label{eq: DP system m2}
		\\
		\partial_t \, (m_1 u_1)  +   \partial_x \, \left(m_1 u_1^2 \right) + \alpha \partial_x p_1  = & M ,
		\label{eq: DP system m1u1}
		\\
		\partial_t \, (m_2 u_2) +  \partial_x \, \left(m_2 u_2^2 \right) + (1-\alpha) \, \partial_x p_2  = & - M ,
		\label{eq: DP system m2u2}
		\\
		\partial_t \, (m_1 h_1)  +   \partial_x \, \left(m_1 h_1 u_1 \right) - \alpha \left( \, \partial_t p_1 + u_1 \partial_x p_1 \right) & =  -u_1  \frac{q_{1i}''}{L_s} ,
		\label{eq: DP system m1h1}
		\\
		\partial_t \, (m_2 h_2) +  \partial_x \, \left(m_2 h_2 u_2 \right) - (1-\alpha) \left( \, \partial_t p_2 + u_2 \partial_x p_2 \right) & = \frac{q_{2i}''}{L_s} ,
		\label{eq: DP system m2h2}
		\\
	\frac{m_2 R \ddot{R}}{(1-\alpha) (1- ikR)} + \frac{m_2 \dot{R}^2}{(1-\alpha)} 
	\left(
	\frac{2}{1-ikR} -\demi - \left( \frac{kR}{1-ikR}\right)^2 
	\right) 
	+
	\frac{4 \mu_2 \dot{R}}{R} 
	\left( 1 - \frac{(kR)^2}{2(1-ikR)}\right) 
	+ \frac{2 \sigma}{R}
	&=p_{1i} - p_2,
	\label{eq: DP system - R}
\end{align}
 \label{eq:DP-system}
\end{subequations}
where $q_{1i}''$ and $q_{2i}''$  are interfacial heat fluxes, $1/L_s$ is
 the interfacial area density and M accounts for interactions like drag force, virtual mass or Basset force (see \citep{cheng85}).
Relation~\eqref{eq: DP system - R} is obtained thanks to \eqref{eq: bernoulli one bubble} and \eqref{eq: Laplace jump relation} and governs the evolution of $R$, by means of $\alpha$ in \eqref{eq: DP system m1}-\eqref{eq: DP system m2h2} it allows to account for small-scale two-phase interface dynamics in the bulk dynamics. 

Finally, the system~\eqref{eq:DP-system} has to be complemented with an evolution equation for the number density $n$. If one supposes that no coalescence nor breakup can occur, $n$ verifies the conservation equation 
$$
\partial_t n + \partial_x (nu_1) = 0.
$$


\section{Eigenstructure of the two-phase model with micro-inertia}
\label{section: appendix eigenstructure} 

We consider the sole convective part of system~\eqref{eq: general consv system + dissipation} for one-dimensional problems by discarding the source terms. 
For smooth solutions, the obtained system may be expressed using the variable
$\bV = (\rho,u,Y,\alpha,w)^T$ as follows
\begin{equation*}
\partial_t
\bV+
A(\bV)
\partial_x
\bV
=0
,\qquad
A=
\begin{pmatrix}
u & \rho & 0 & 0 &0 
\\
\frac{c^2}{\rho}
& 
u
&
\frac{1}{\rho}\frac{\partial p}{\partial Y}  + Y \rho w^2
&
\frac{1}{\rho}\frac{\partial p}{\partial \alpha} 
&
w \rho Y^2
\\
0 & 0 & u & 0 & 0
\\
0 & 0 & 0 & u & 0
\\
0 & 0 & 0 & 0 & u
\end{pmatrix}
.
\end{equation*}
The matrix $A(\bV)$ possesses three distinct eigenvalues: $u\pm c$ and $u$ associated respectively with the eigenvectors
$$
\mathbf{R}_{u\pm c}=
(\rho, \pm c, 0, 0, 0)^T
,
\mathbf{R}^{(1)}_{u}=
\left(\frac{\partial p}{\partial Y}  + Y \rho w^2, 0, -c^2 , 0, 0\right)^T
,
\mathbf{R}^{(2)}_{u}=
\left(\frac{\partial p}{\partial \alpha} , 0, 0, -c^2 , 0\right)^T
,
\mathbf{R}^{(3)}_{u}=
(\rho ^2 w Y^2 , 0, 0 , 0, -c^2 )^T
.
$$

\end{appendices}

\section*{Acknowledgments}
The Ph.D. of F.~Drui is funded by a CEA/DGA (Direction G\'en\'erale de l'Armement - French Department of Defense) grant.  
 
\bibliographystyle{plainnat}

\end{document}